%% file: article_arxiv.tex
\documentclass[USenglish, onecolumn]{scrartcl}
\usepackage[utf8]{inputenc}
\usepackage[USenglish]{babel}
\usepackage{csquotes}
\usepackage[a4paper, margin=2cm, bottom=3cm]{geometry}

\usepackage{lmodern}
\usepackage{microtype}

\usepackage{authblk}

\usepackage{abstract}
\newcommand{\placeabstract}[1]{\begin{abstract}\noindent{}#1\end{abstract}\twocolumn}

\newcommand{\qt}[1]{`#1'}

\newcommand{\dir}{./}
\newcommand{\ldir}{\dir literature.bib}
\newcommand{\gdir}{\dir Graphics/}

\usepackage[
    backend=biber,
    style=numeric-comp,
  	maxalphanames=1,
    sortlocale=en_US,
    natbib=true,
    url=false, 
    doi=true,
    eprint=true,
    giveninits=true,
  	maxbibnames=99,
    maxcitenames=1,
    sorting=none
]{biblatex}

\DefineBibliographyStrings{USenglish}{
  andothers = \textit{et al.}
}
\renewbibmacro*{volume+number+eid}{
  \printfield{volume}\iffieldundef{volume}{}{\mkbibbold{\addcomma}}
  \thefield{number}
}
\renewbibmacro*{journal+issuetitle}{%
  \usebibmacro{journal}%
  \setunit*{\addspace}%
  \iffieldundef{series}
    {}
    {\newunit
     \printfield{series}%
     \setunit{\addspace}}%
  \usebibmacro{volume+number+eid}%
  \newunit%
}
\renewbibmacro*{publisher+location+date}{%
  \printlist{location}%
  \iflistundef{publisher}
    {\setunit*{\addcomma\space}}
    {\setunit*{\addcolon\space}}%
  \printlist{publisher}%
  \setunit*{\addspace}%
  \usebibmacro{date}%
  \newunit%
}
\DeclareBibliographyDriver{article}{%
  \usebibmacro{bibindex}%
  \usebibmacro{begentry}%
  \usebibmacro{author/translator+others}%
  \printlist{language}%
  \newunit\newblock
  \usebibmacro{byauthor}%
  \newunit\newblock
  \usebibmacro{bytranslator+others}%
  \newunit\newblock
  \printfield{version}%
  \newunit\newblock
  \usebibmacro{in:}%
  \usebibmacro{journal+issuetitle}%
  \newunit
  \usebibmacro{byeditor+others}%
  \newunit
  \usebibmacro{note+pages}%
  \setunit{\addspace}%
  \usebibmacro{issue+date}%
  \setunit{\addcolon\space}%
  \usebibmacro{issue}%
  \newunit\newblock
  \usebibmacro{doi+eprint+url}%
  \newunit\newblock
  \usebibmacro{addendum+pubstate}%
  \setunit{\bibpagerefpunct}\newblock
  \usebibmacro{pageref}%
  \usebibmacro{finentry}}
\renewbibmacro{in:}{%
  \ifentrytype{article}{}{\printtext{\bibstring{in}\intitlepunct}}%
}

\AtEveryBibitem{\clearfield{month}}
\AtEveryBibitem{\clearfield{day}}
\AtEveryBibitem{%
  \ifentrytype{misc}{%
  }{%
    \clearfield{eprint}%
  }%
}
\DeclareFieldFormat[article]{title}{{#1}}
\DeclareFieldFormat[inproceedings]{title}{\qt{#1}}
\DeclareFieldFormat[inproceedings]{date}{\mkbibparens{#1}}
\DeclareFieldFormat[misc]{title}{{#1}}
\DeclareFieldFormat[article]{volume}{\mkbibbold{#1}}
\DeclareFieldFormat[article]{number}{\mkbibparens{#1}}
\DeclareFieldFormat{pages}{#1}
\usepackage{xcolor}
\usepackage{soul}

\newboolean{highlightdiff}
\setboolean{highlightdiff}{false}
\ifthenelse{\boolean{highlightdiff}}{%
	\DeclareRobustCommand{\hlrm}[1]{{\textcolor{red}{\setstcolor{red}\st{#1}}}}%
	\DeclareRobustCommand{\hlrmm}[1]{{\textcolor{red}{\setstcolor{red}\st{\mbox{#1}}}}}%
}{%
	\DeclareRobustCommand{\hlrm}[1]{}%
	\DeclareRobustCommand{\hlrmm}[1]{}%
}

\usepackage{amsmath}
\usepackage{amssymb}
\usepackage{bm}
\usepackage{nicefrac}
\usepackage{mathtools}
\usepackage{siunitx}
\usepackage{algorithm}
\usepackage{algpseudocode}

\usepackage{xr}
\usepackage{hyperref}

\newboolean{OnlyCaptions}
\setboolean{OnlyCaptions}{false}
\newboolean{NoCaptions}
\setboolean{NoCaptions}{false}

\usepackage{graphicx}
\ifthenelse{\not\boolean{NoCaptions}}{%
	\usepackage[labelfont=bf, format=plain, labelsep=period]{caption}%
}{%
	\renewcommand{\caption}[1]{}%
}
\addto\extrasUSenglish{%
}
\newcommand{\subfignum}[1]{#1}
\newcommand{\subfiglabel}[1]{\textbf{\subfignum{#1}}}
\newcommand{\subfigref}[2]{\autoref{#1}\subfignum{#2}}

\NewDocumentCommand{\fig}{s O{1} O{} O{} O{} O{figure} m}{%
	\ifthenelse{\not\boolean{OnlyCaptions}}{%
		\begin{\IfBooleanTF #1{#6*}{#6}}[#4]%
			\centering%
			\includegraphics[width = #2\textwidth]{\gdir #7}%
			\caption[#5]{#3}%
			\label{fig:#7}%
		\end{\IfBooleanTF #1{#6*}{#6}}%
	}{#3}%
}

\begin{document}	

\input{header.tex}
\maketitle

\input{content.tex}

\onecolumn
\printbibliography
\end{document}


\input{header.tex}
\subtitle{Supplementary Information}
\maketitle

\section{Sensor design}
\label{s:sensor-design}

\subsection{Mechanical setup}
\label{ss:machanical-setup}
\begin{figure}
	\centering
	\includegraphics[scale=.3]{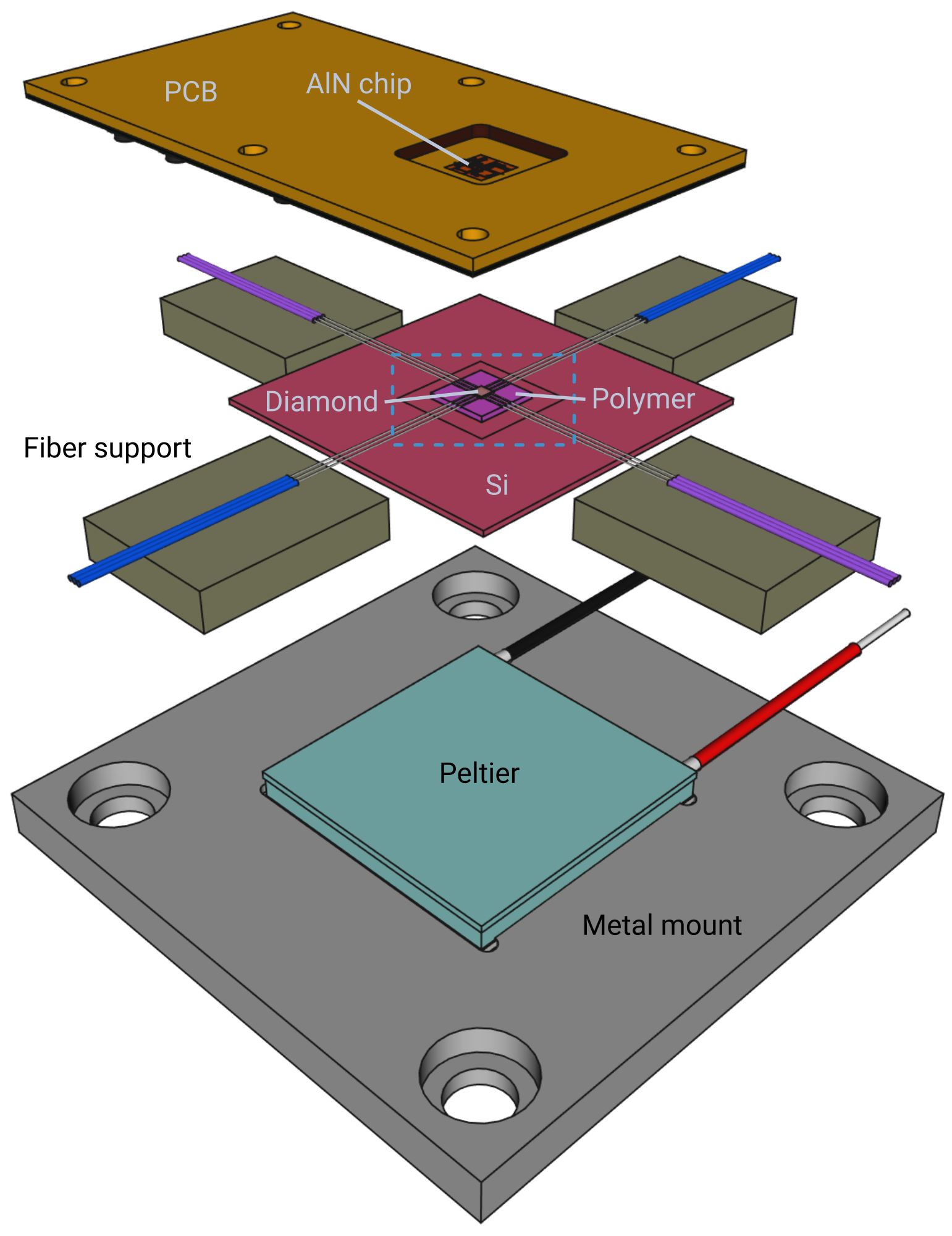}
	\caption{\textbf{Multi-pixel sensor setup.}
	The diamond substrate and fibers are embedded into the PolyBoard platform (Polymer) located on top of a silicon (Si) submount (\autoref{ss:polymer-board}).
	Microwave inductors fabricated on an aluminum nitride (AlN) chip embedded into a standard printed circuit board (PCB) above the diamond substrate deliver microwaves for infrared absorption ODMR measurements (\autoref{ss:microwave-inductor}).
	Four fiber supports avoid bending of the optical fibers close to the sensor region.
	A Peltier device in between the silicon submount and the metal mount stabilizes the sensor's temperature.
	We detail the sensor's main parts (dashed rectangle) in the main manuscript.
	}
	\label{fig:setup_annotated}
\end{figure}

\autoref{fig:setup_annotated} illustrates the multi-pixel sensor's overall mechanical setup.
The diamond substrate and the fibers to optically access the pixels are at the core of our sensor.
They are integrated into the Fraunhofer Heinrich-Hertz-Institut's PolyBoard platform (\autoref{ss:polymer-board}), which is stacked on top of a silicon submount, a Peltier device for temperature control and a metal mount.
A printed circuit board (PCB) above the diamond substrate hosts microwave inductors to drive the NV$^-$'s spin transitions for infrared absorption ODMR (\autoref{ss:microwave-inductor}).
Three xyz positioner stages and a rotational stage allows to adjust the PCB's position and its angle around the vertical axis relative to the sensor stack.

\subsection{Diamond substrate}
\label{ss:diamond-substrate}
\begin{figure}
	\centering
	\includegraphics[scale=.7]{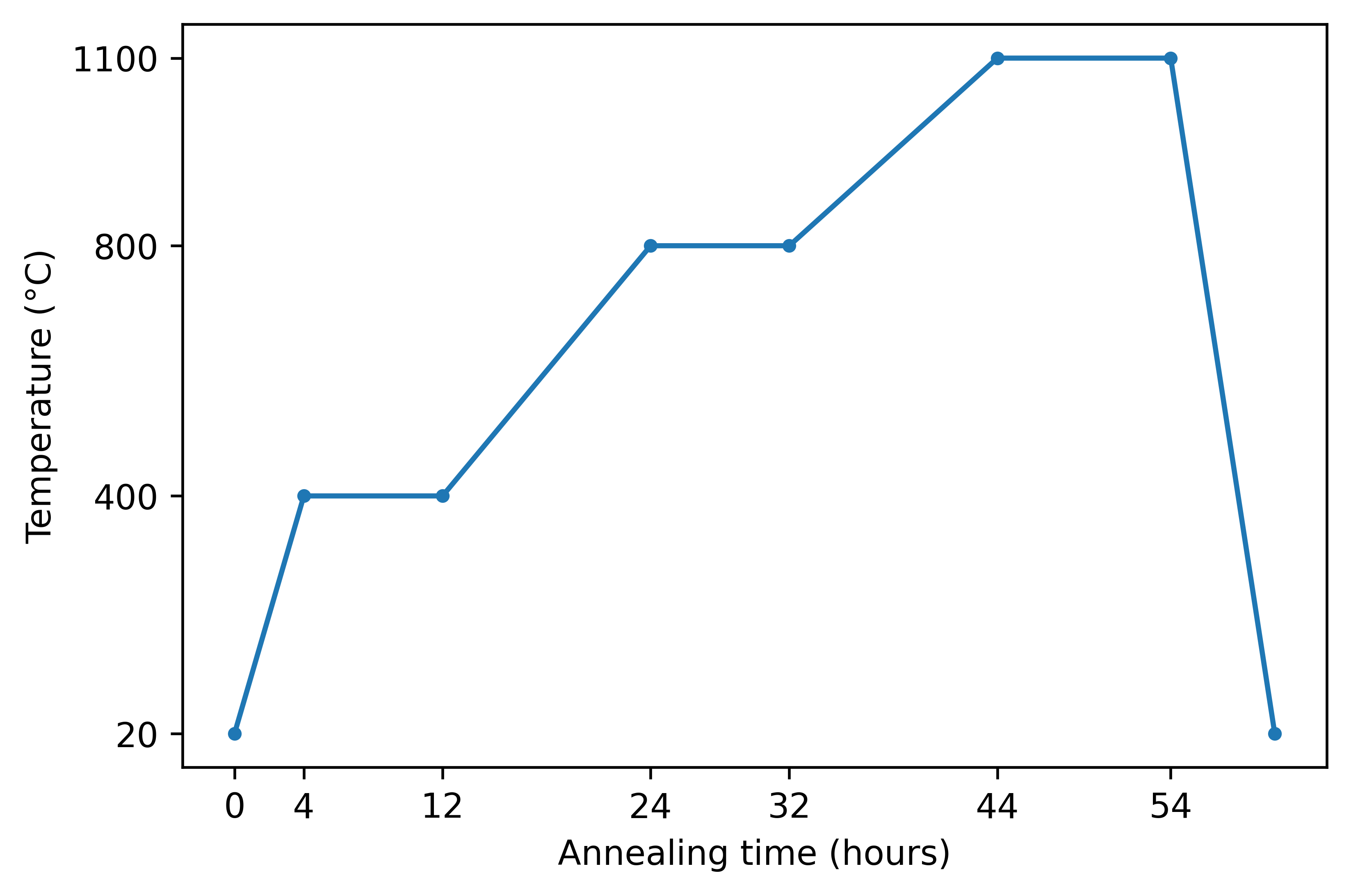}
	\caption{\textbf{Diamond substrate annealing temperature profile.}}
	\label{fig:temp_curve}
\end{figure}

We increase the NV center density of our two diamond substrates by electron irradiation followed by an annealing step.
Firstly, the diamond substrate A (B) is irradiated with a $\SI{7}{\mega\electronvolt}$ electron beam and a fluence of $\SI{1.5e17}{\centi\meter^{-2}}$ $\left(\SI{0.5e17}{\centi\meter^{-2}}\right)$ at the Leibniz Institute of Surface Engineering.
Secondly, we anneal the substrates in vacuum according to the temperature profile displayed in \autoref{fig:temp_curve} at the Walter Schottky Institute, Technical University of Munich.
To efficiently couple pump light into the diamond substrates and to minimize losses when measuring the infrared transmission across them, their four vertical side facets are polished to a roughness lower than $\SI{3}{\nano\meter}\,\mathrm{Ra}$ by Applied Diamond, Inc.
Finally, we clean the diamond substrates with a boiling mixture of nitric, perchloric, and sulfuric acid (HNO$_3$:H$_2$SO$_4$:HClO$_4$).

\subsection{Polymer board and assembly}
\label{ss:polymer-board}
The polymer optical interposer was manufactured in the PolyBoard platform of Fraunhofer Heinrich-Hertz-Institut \citep{Kleinert2021}.
A 4-inch silicon wafer acts as a mechanical substrate for the polymer layers.
After applying an adhesion promoter (Exfix ZAP-1020 by ChemOptics Inc.) to the wafer surface, a layer of a fluorinated acrylate (ZPU12-RI by ChemOptics Inc.) with a thickness of approx. $\SI{80}{\micro\meter}$ is deposited with spin-coating and subsequently UV-cured in a nitrogen atmosphere.
Next, a hard bake takes place followed by the application of a second polymer layer with a thickness of approx. $\SI{30}{\micro\meter}$.
The second polymer layer is cured in the same way as the first layer.
Electron beam evaporation now deposits titanium on the wafer surface before a positive-type photoresist is spin-coated.
Contact lithography on a mask aligner (SÜSS MicroTec MA8) is used to expose the photo resist with a mask containing the U grooves and free-space regions of the polymer interposer.
We structure the underlying titanium layer by means of the developed photo resist, which is subsequently removed.
During dry etching of the polymer with oxygen-based inductively coupled plasma reactive ion etching (ICP-RIE), the structured titanium acts as a hard mask.
Laser interferometry controls the etch depth in-situ.
After dry etching, the titanium mask is removed and the wafer is separated into individual chips by dicing.

Graded-index (GRIN) lenses in the polymer interposer act as collimating and focusing elements.
These GRIN lenses are fabricated by polishing a graded-index fiber (Yangtze Optical Fibre and Cable) with a numerical aperture of $\num{0.14}$ to a length of $\SI{840}{\micro\meter}$, corresponding to a GRIN lens pitch of $\num{0.25}$.
The GRIN lenses' outer diameter of $\SI{125}{\micro\meter}$ matches the width of the U groove sections on the polymer interposer.

In an optical assembly step, we insert GRIN lenses, optical fibers and the diamond substrate into the respective openings in the polymer interposer.
First, the three pump light paths are set up one after another by inserting GRIN lenses and single-mode fiber pigtails (SM450, FC/APC connector on one side, cleaved on the other side) into the U grooves such that they are in contact.
Second, we apply UV-curable epoxy on both sides to provide mechanical fixation and refractive index matching between the fibers and GRIN lenses.
After setting up the three pump paths, the three perpendicular infrared light paths are assembled similarly employing suitable fibers (HI1060-J9, FC/APC connector on one side, cleaved on the other side).
Then, we insert the diamond substrate into the on-chip free-space section.
Its rotational alignment is optimized actively by monitoring the transmission through the optical paths before UV-curable epoxy is applied.
Finally, we place the optical assembly onto a silicon submount, which provides strain relief for the optical fibers.

\subsection{Microwave inductor}
\label{ss:microwave-inductor}
\begin{figure}
	\centering
	\includegraphics[scale=.7]{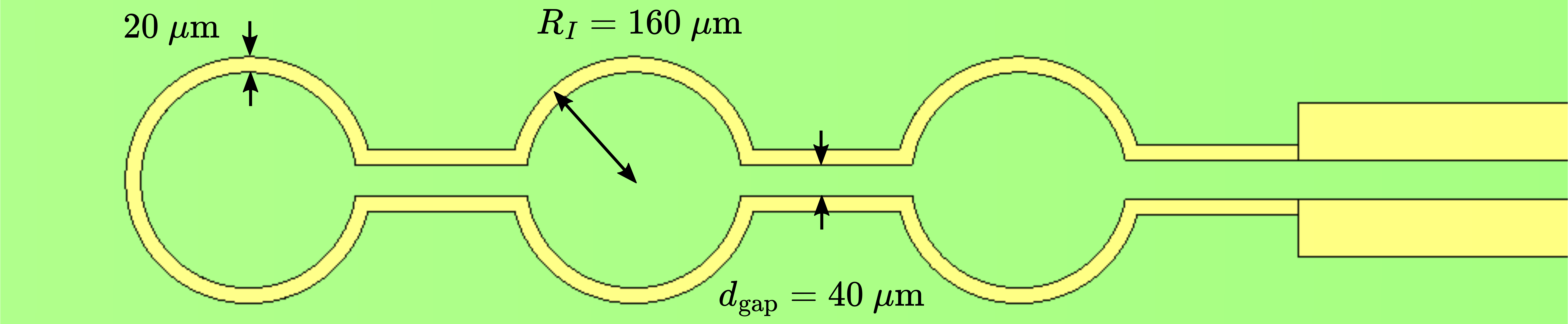}
	\caption{\textbf{Microwave inductor line consisting of three connected rings.}
	The AlN chip contains three such lines.
	Thus, each sensor pixel is located underneath one ring after alignment with respect to the sensor stack (\autoref{ss:machanical-setup}).}
	\label{fig:RFInductorTop}
\end{figure}

The microwave inductors aim to generate a magnetic flux density of approximately $\SI{1}{\milli\tesla}$ at an operating frequency of $f_\mathrm{I,op}=\SI{2.88(5)}{\giga\hertz}$ within the pixel volumes.
Due to fabrication constraints and to ensure that the inductors' first resonance frequency $f_\mathrm{I,res}$ is much larger than their operating frequency $f_\mathrm{I,op}$, we chose a single winding design.
Each winding possesses a radius $R_\mathrm{I}=\SI{160}{\micro\meter}$, a width of $\SI{20}{\micro\meter}$, and a thickness of $\SI{5}{\micro\meter}$.
Based on the Biot-Savart law, the absolute value of the microwave magnetic flux density $B_\mathrm{I}$ at the pixels' centers normalized with respect to the current $I_\mathrm{I}$ flowing through the inductor is approximated by
\begin{align}
\label{eq:normfluxdens}
K_\mathrm{I}=\frac{B_\mathrm{I}}{I_\mathrm{I}} \approx \frac{2\mathrm{\pi} R_\mathrm{I} - 2d_\mathrm{gap}}{2\mathrm{\pi} R_\mathrm{I}}\frac{\mathrm{\mu}_0 R_\mathrm{I}^2}{2\sqrt{R_\mathrm{I}^2 + d_\mathrm{pix}^2}^{3}}\,,
\end{align}
where $d_\mathrm{gap}=\SI{40}{\micro\meter}$ is the distance between the feeding lines of the inductor (refer to \autoref{fig:RFInductorTop}), $d_\mathrm{pix} = \SI{150}{\micro\meter}$ the distance between the inductor and the respective pixel underneath, and $\mu_0 $ the vacuum magnetic permeability.
\autoref{eq:normfluxdens} yields $K_\mathrm{I} \approx \SI{1.40}{\milli\tesla\per\ampere}$ for the microwave inductor geometry as depicted in \autoref{fig:RFInductorTop}.
The analytic result agrees well with $K_\mathrm{I} \approx \SI{1.33}{\milli\tesla\per\ampere}$ obtained numerically with the magnetostatic solver of CST Studio Suite.
Both values suggest that an inductor current lower than $\SI{1}{\ampere}$ allows for the desired magnetic flux density in the order of $\SI{1}{\milli\tesla}$.

\begin{figure}
	\centering
	\includegraphics[scale=.8]{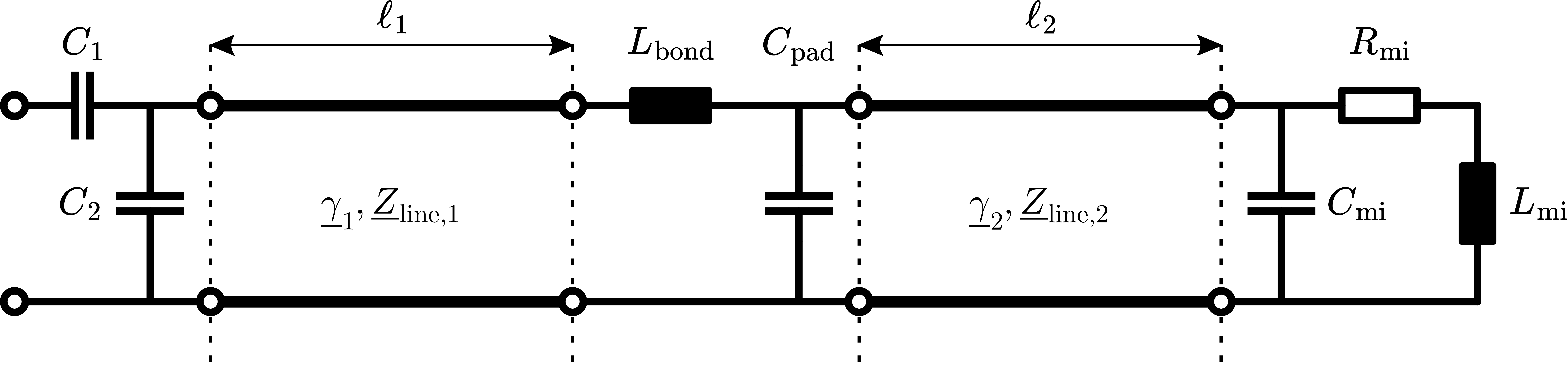}
	\caption{%
	\textbf{Equivalent circuit diagram of the microwave inductor setup.}
	The microwave inductor is modeled by the inductance $L_\mathrm{mi}$, the resistor $R_\mathrm{mi}$, and the capacitor $C_\mathrm{mi}$.	
	The transmission line characterized by the propagation constant $\underline{\gamma}_2$, the characteristic impedance $\underline{Z}_{\mathrm{line},2}$, and the length $\ell_2$ describes the microstrip line on the AlN chip connecting the microwave inductor with the landing bond pad.
	$C_\mathrm{pad}$ accounts for the influence of the landing pad, while $L_\mathrm{bond}$ takes the influence of the bond wire connecting the PCB with the AlN chip into account.
	The transmission line characterized by $\underline{\gamma}_1$, $\underline{Z}_{\mathrm{line},1}$, and $\ell_1$ models the microstrip line on the PCB.
	Impedance matching is provided by the capacitors $C_1$ and $C_2$. 
	}
	\label{fig:eq_model_mw_inductor}
\end{figure}

Fabricating the microwave inductors onto an aluminum nitride (AlN) chip embedded into an outer standard printed circuit board (PCB) prevents excessive heating.
\autoref{fig:eq_model_mw_inductor} presents an equivalent circuit diagram of the entire microwave inductor setup consisting of the AlN chip and the PCB.
The inductance $L_\mathrm{bond}=\SI{333}{\pico\henry}$ models the bond wire which connects the outer PCB with the AlN chip.
The elements on the left side of this inductance represent the circuit on the PCB, whereas the elements on the right side correspond to the circuit on the AlN chip.
On the AlN chip, an inductance $L_\mathrm{mi} = \SI{2.6}{\nano\henry}$ with a resistor $R_\mathrm{mi} = \SI{2}{\ohm}$ in parallel to a capacitor $C_\mathrm{mi} = \SI{398}{\femto\farad}$ describe the microwave inductors.
Thus, their first resonance frequency becomes $f_\mathrm{I,res} \approx 1/(2\mathrm{\pi}\sqrt{C_\mathrm{mi}L_\mathrm{mi}})=\SI{4.9}{\giga\hertz}$, which is much larger than the operating frequency $f_\mathrm{I,op}$.
The connection between the microwave inductor and the bond pad on the AlN chip is described by a transmission line with a propagation constant $\underline{\gamma}_2$, a characteristic impedance $\underline{Z}_{\mathrm{line},2}$, and a length $\ell_2 = \SI{5425}{\micro\meter}$.
$C_\mathrm{pad}=\SI{107}{\femto\farad}$ takes the bond pad's influence on the AlN chip into account.
Similarly, a transmission line with a propagation constant $\underline{\gamma}_1$, a characteristic impedance $\underline{Z}_{\mathrm{line},1}$, and a length $\ell_1 = \SI{20.5}{\milli\meter}$ models the microstrip line on the PCB.
The capacitors $C_1$ and $C_2$ provide impedance matching to $\SI{50}{\ohm}$ to minimize the reflection coefficient at the PCB's input terminals.
All propagation constants and characteristic impedances of the lines are frequency-dependent complex values.
They are determined using a microstrip line model \citep{Kobayashi1988} applied to the cross sections of the AlN chip and the PCB, respectively.
We employ CST Studio Suite's frequency-domain solver to analyze the microwave inductor's properties around the operating frequency $f_\mathrm{I,op}$.

\section{Magnetic field camera ODMR measurements}
\label{s:odmr-measurements}

\subsection{Pixel integrity}
\label{ss:pixel-integrity}
\begin{figure*}
	\centering
	\includegraphics[scale=1]{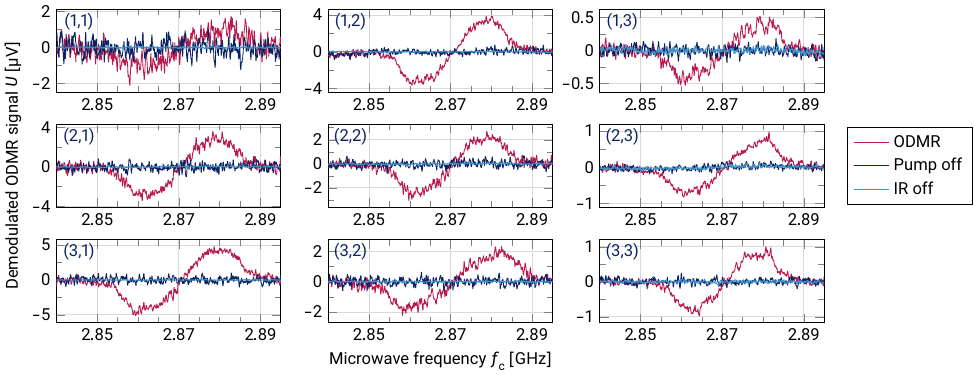}
	\caption{\textbf{IRA-ODMR signatures (red curves) of the multi-pixel magnetic field camera's pixels $\left( i,j\right)$.}
	There is no offset magnetic field applied.
	For the dark (light) blue curves, the pump (infrared) laser is turned off.
	}
	\label{fig:Pixels_ODMR}
\end{figure*}

To prove the integrity of the magnetic field camera's pixels, we record their ODMR signatures without an offset magnetic field.
Moreover, we compare the ODMR-active configuration with the pump as well as the infrared laser turned on to configurations with one of the lasers tuned off.
\autoref{fig:Pixels_ODMR} displays the respective results.
Clearly, the ODMR signatures (red curves) vanish if either the pump or the infrared laser is turned off.
This signifies clear evidence that the recorded ODMR signals originate from infrared absorption instead of residual red fluorescence light coupled into the infrared fibers.
Due to an insufficient beam overlap between the pump and the infrared beams, the signal-to-noise ratio of the ODMR signature related to pixel $\left(\num{1},\num{1}\right)$ does not suffice to extract magnetic fields in the following.
Thus, this pixel is excluded from further analysis.

\subsection{Magnetic field measurements}
\label{ss:camera-B-measurements}
\begin{figure*}
	\centering
	\includegraphics[scale=1]{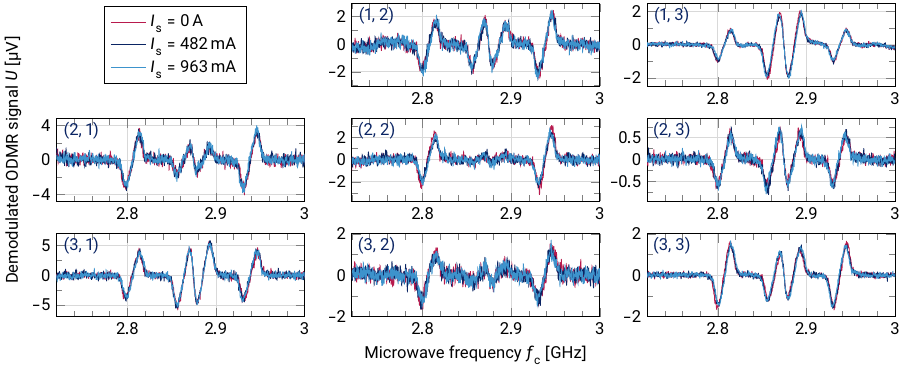}
	\caption{\textbf{Raw IRA-ODMR signatures recored with the magnetic field camera for imaging a solenoid coil's magnetic field.}
	A current $I_\mathrm{s}$ drives the solenoid coil located above pixel $\left(\num{3},\num{3}\right)$.
	The red curves show ODMR signatures without current $\left( I_\mathrm{s}=\SI{0}{\ampere}\right)$.
	Their lower (upper) resonance possesses a slightly lower (higher) resonance frequency compared to the higher current cases (blue curves).
	Frequency shifts become visible with linear fits around the resonances.
	Current-dependent frequency shifts are exploited for magnetic field sensing.
	}
	\label{fig:Camera_ODMR}
\end{figure*}

Applying an offset magnetic field aligned along the $\left[0 \bar{1} 1\right]$-direction splits the distinct NV subensembles' resonances, and thus allows for magnetic field measurements.
We record an image of the magnetic field produced by a current $I_\mathrm{s}$-driven solenoid coil located above pixel $\left(\num{3},\num{3}\right)$.
\autoref{fig:Camera_ODMR} shows the pixels' ODMR signatures for three different solenoid currents.
Solenoid current-dependent frequency shifts are hardly visible in the raw ODMR data, but become apparent by fitting the resonances with linear functions to extract the resonance frequencies $f_-$ and $f_+$ (refer to main manuscript).

\begin{figure}
	\centering
	\includegraphics[scale=1]{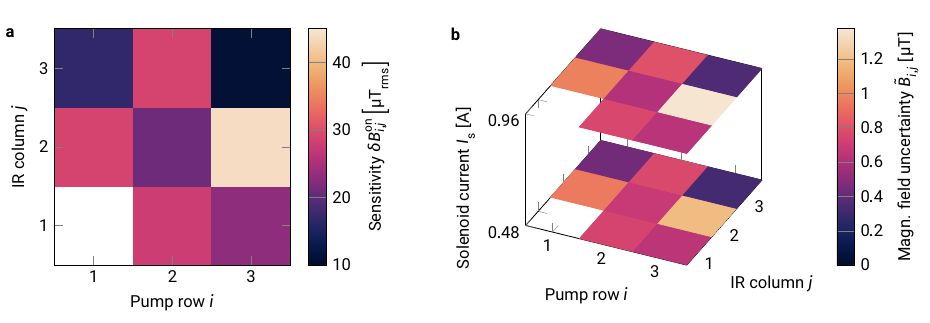}
	\caption{\textbf{Magnetic field camera pixel sensitivity and magnetic field uncertainty}
	\subfiglabel{a} Magnetic sensitive noise amplitude spectral densities $S^\mathrm{on}_{i,j}$ integrated up to the lock-in cutoff frequency.
	\subfiglabel{b} Magnetic field uncertainties extracted from the $\num{95}\,\%$ confidence bounds of the linear fits around the resonances displayed in \autoref{fig:Camera_ODMR}.
	}
	\label{fig:Camera_Unc}
\end{figure}

Along with an image of the magnetic field, we also extract the sensitivity for each pixel by integrating the pixels' magnetic sensitive noise amplitude spectral densities $S^\mathrm{on}_{i,j}$ up to the lock-in cutoff frequency (\subfigref{fig:Camera_Unc}{a}).
Furthermore, the measured per-pixel magnetic field uncertainty is given by
\begin{align}
\tilde{B}_{i,j} = \frac{\sqrt{\tilde{f}_{i,j,+}\!\left(I_\mathrm{s}\right)^2 + \tilde{f}_{i,j,-}\!\left(I_\mathrm{s}\right)^2 + \tilde{f}_{i,j,+}\!\left(0\right)^2 + \tilde{f}_{i,j,-}\!\left(0\right)^2}}{2\gamma}\,,
\end{align}
where the resonance frequency uncertainties $\tilde{f}_{i,j,\pm}\!\left(I_\mathrm{s}\right)$ originate from the $\num{95}\,\%$ confidence bounds of the linear fits around the resonances.
For each pixel, the resulting magnetic field uncertainty is lower than $\SI{1.4}{\micro\tesla}$ (\subfigref{fig:Camera_Unc}{b}).
Since the measured absolute magnetic fields exceed their uncertainty by at least one order of magnitude, they are hardly affected by fit uncertainties.

\subsection{Reconstruction of solenoid position}
\label{ss:solenoid-position}
The recorded image of the solenoid's magnetic field allows the reconstruction of its spatial position relative to the magnetic field camera.
This reconstruction is based on a minimization of the difference between the absolute values of the per-pixel measured magnetic field $B_{i,j}$ and the solenoid's simulated magnetic field $B_\mathrm{s}\!\left(x,y,z\right)$.
We employ the Python package \textit{Magpylib} \citep{Ortner2020} to estimate the solenoid's magnetic field.
To obtain the solenoid position $\left(x, y, z\right)$ as the vector difference between the solenoid's center and the center pixel $\left(\num{2},\num{2}\right)$ as well as the corresponding position uncertainty $\left(\delta x,\delta y,\delta z\right)$, we first draw about $\num{11000}$ samples $B^{\left( m\right)} \coloneq \left\lbrace B^{\left( m\right)}_{i,j}\right\rbrace$ from a multivariate normal distribution with a mean vector $\bm{\mu}=\left(B_{1,1},\dots,B_{3,3}\right)^\intercal$ and a covariance matrix $\bm{\Sigma}=\mathrm{diag}\!\left(\tilde{B}_{1,1}^2,\dots,\tilde{B}_{3,3}^2\right)$ by Monte Carlo sampling.
For each sample $B^{\left( m\right)}$, we perform the minimization
\begin{align}
\min_{x,y,z} \sqrt{\sum^3_{i,j=1}\,\sum^{\left(v-1\right)/2}_{k,l=-\left(v-1\right)/2}\left| B^{\left( m\right)}_{i,j} - B_\mathrm{s}\!\left( x+d\left( i-2\right) +\frac{w k}{v-1},\: y+d\left(j-2\right) +\frac{w l}{v-1},\: z\right)\cos\left(\alpha\right)\right|^2}
\end{align}
with a Nelder-Mead algorithm, where $d=\SI{500}{\micro\meter}$ is the pixel lattice constant and $w=\SI{80}{\micro\meter}$ the pixel size.
The angle $\alpha=\SI{35.3}{\degree}$ takes the orientation of the $\left[\bar{1} 1 \bar{1}\right]$- and $\left[\bar{1} \bar{1} 1\right]$-oriented NV subensembles used for sensing into account.
$i,j$ sum over all pixels and $k,l$ sample the simulated field $v\times v$ times within each pixel.
$v$ has to be odd and $\geq\num{3}$.
We choose $v=\num{3}$.
Building histograms for each direction from the resulting position samples $\left(x^{\left( m\right)}, y^{\left( m\right)}, z^{\left( m\right)}\right)$ and fitting the histograms with normal distributions yields the solenoid position and its uncertainty as the histograms' mean values and standard deviations, respectively.

\section{Noise amplitude spectral density}
\label{s:noise-asd}

\subsection{Multi-pixel sensor ASD measurements}
\label{ss:asd-measurements}
We measure magnetic sensitive, insensitive and electronic noise amplitude spectral densities (ASD) $S\!\left(f\right)$ according to the following procedure.
First, we record a $\SI{100}{\second}$ long time series of the demodulated ODMR signal $U\!\left(t\right)$ at a fixed microwave center frequency $f_\mathrm{c}$ and transform it into a time-dependent magnetic field
\begin{align}
B\!\left(t\right) = \frac{1}{a_+ \gamma}\left(U\!\left(t\right) - \underset{t}{\mathrm{mean}}\!\left[U\!\left(t\right)\right]\right)
\end{align}
knowing the slope $a_+$ of the upper resonance's linear fit.
Second, a Fourier transform and subsequent normalization by the frequency spacing $\Delta f$ between the Fourier-transformed samples yields the single-sided ASD as
\begin{align}
S\!\left(f\right) = \frac{2\left|\mathcal{F}\!\left[B\!\left(t\right)\right]\right|}{\sqrt{\Delta f}}\,.
\end{align}
The rms noise $\delta B_\mathrm{rms}$ corresponding to a frequency interval $\left[f_1,f_2\right]$ is the integral of the ASD
\begin{align}
\delta B_\mathrm{rms} = \int_{f_1}^{f_2}S\!\left(f\right)\,\mathrm{d}f\,.
\end{align}

\subsection{Infrared shot noise}
\label{ss:shot-noise}
The amplitude spectral density of optical shot noise caused by infrared photons with a wavelength of $\lambda_\mathrm{s}=\SI{1042}{\nano\meter}$ impinging on the two photodiodes of the balanced photodetector (PDB450C from Thorlabs with $\times 10^5$ difference output gain) can be inferred directly as
\begin{align}
\label{eq:S_SN}
s^\mathrm{SN} = \sqrt{2}\sqrt{2 h \frac{c}{\lambda_\mathrm{s}} \bar{P}_\mathrm{IR}}
\end{align}
by measuring the average infrared optical power $\bar{P}_\mathrm{IR}$ at the detector.
In \autoref{eq:S_SN}, the first factor of $\sqrt{2}$ takes into account that \textit{two} photodiodes become involved in the balanced detection.
$h$ is the Planck constant and $c$ the speed of light.
The average infrared power is derived from the average voltages measured at the PDB450C's monitor outputs $\bar{U}_\mathrm{Mon}$, the monitor output's conversion gain $g_\mathrm{Mon}=\SI{10}{\volt\per\milli\watt}$, and the detectors efficiency $R=\num{68.3}\,\%$ at $\lambda_\mathrm{s}$.
Thus, we obtain
\begin{align}
\bar{P}_\mathrm{IR} = \frac{\bar{U}_\mathrm{Mon}}{g_\mathrm{Mon} R}\,.
\end{align}
Converting to units of the magnetic field measured at the upper resonance with slope $a_+$ yields
\begin{align}
\label{eq:S_SN_B}
S^\mathrm{SN} = s^\mathrm{SN} \times\frac{g_\mathrm{Sig} R}{a_+ \gamma}\,,
\end{align}
with the difference signal output's conversion gain $g_\mathrm{Sig}=\SI{50}{\volt\per\milli\watt}$.

\section{ODMR contrast}
\label{s:odmr-contrast}

\subsection{8-level NV model}
\label{ss:8-level-model}
To simulate the ODMR contrast $C$ of our measurements, we apply the approach detailed in \citep{Dumeige2013} and extended to an 8-level NV model including its neutral charge state in \citep{Dumeige2019}.
The labeling of transition rates and related quantities equals the labeling introduced in \citep{Dumeige2019}.
With the NV density $n$ and the initial state occupation $\mathcal{N}_0 = \left(0,0,0,0,0,n,0,0\right)^\intercal$, the occupation densities of each level $\mathcal{N} = \left(n_1,n_2,n_3,n_4,n_5,n_6,n_7,n_8\right)^\intercal$ after a propagation distance $z$ become $\mathcal{N}\!\left(z\right) = \mathcal{M}^{-1}\!\left(z\right) \mathcal{N}_0$, with the matrix

\scalebox{.79}{\parbox{\linewidth}{%
\begin{align}
\mathcal{M} &= \begin{pmatrix}
-\left[W_\mathrm{g}+W_\mathrm{MW}\right] & W_\mathrm{MW} & k_{31} & 0 & 0 & k_{61} & 0 & W_\mathrm{r}/2\\
W_\mathrm{MW} & -\left[W_\mathrm{g}+W_\mathrm{MW}\right] & 0 & k_{42} & 0 & k_{62} & 0 & W_\mathrm{r}/2\\
W_\mathrm{g} & 0 & -\left[ k_{31}+k_{35}+W_\mathrm{i}\right] & 0 & 0 & 0 & 0 & 0\\
0 & W_\mathrm{g} & 0 & -\left[ k_{42}+k_{45}+W_\mathrm{i}\right] & 0 & 0 & 0 & 0\\
0 & 0 & k_{35} & k_{45} & -\left[ k_{56} + W_\mathrm{s}\right] & W_\mathrm{s} & 0 & 0\\
1 & 1 & 1 & 1 & 1 & 1 & 1 & 1\\
0 & 0 & W_\mathrm{i} & W_\mathrm{i} & 0 & 0 & -W_\mathrm{g0} & k_{87} \\
0 & 0 & 0 & 0 & 0 & 0 & W_\mathrm{g0} & -\left[ k_{87} + W_\mathrm{r}\right]
\end{pmatrix}\,.
\end{align}
}}

\noindent Here, $W_\mathrm{g}$ $\left( W_\mathrm{g}\right)$ denote the pumping rate of the NV$^-$ (NV$^0$), $W_\mathrm{i}$ $\left( W_\mathrm{r}\right)$ the ionization (recombination) rate from (to) the NV$^-$, $W_\mathrm{s}$ the rate to cycle between the NV$^-$ singlet states, and $W_\mathrm{MW}$ the rate to drive spin-flip transitions with microwaves.
$\Omega_\mathrm{R}$ is the Rabi frequency and $T_2^*$ the electron spin dephasing time associated with $W_\mathrm{MW}$.
$\sigma$ signifies the respective absorption cross section, $I_\mathrm{g}$ $\left( I_\mathrm{s}\right)$ the pump (infrared) intensity, $\lambda_\mathrm{g}$ $\left( \lambda_\mathrm{s}\right)$ the pump (infrared) wavelength, and $k_{ab}$ transition rates from level $a$ to level $b$ (\autoref{t:parameters}).
With the expressions

\begin{align}
W_\mathrm{g} &= \frac{\sigma_\mathrm{g} I_\mathrm{g} \lambda_\mathrm{g}}{h c}\,,\\
W_\mathrm{g0} &= \frac{\sigma_\mathrm{g0} I_\mathrm{g} \lambda_\mathrm{g}}{h c}\,,\\
W_\mathrm{i} &= \frac{\sigma_\mathrm{i} I_\mathrm{g} \lambda_\mathrm{g}}{h c}\,,\\
W_\mathrm{r} &= \frac{\sigma_\mathrm{r} I_\mathrm{g} \lambda_\mathrm{g}}{h c}\,,\\
W_\mathrm{s} &= \frac{\sigma_\mathrm{s} I_\mathrm{s} \lambda_\mathrm{s}}{h c}\,,\\
W_\mathrm{MW} &= \frac{1}{2}\Omega_\mathrm{R}^2 T_2^*\,,
\end{align}

\noindent the system of differential equations

\begin{align}
\begin{dcases}
\frac{\mathrm{d}I_\mathrm{g}}{\mathrm{d}z} = -\left[\sigma_g\left(n_1+n_2\right)+\sigma_\mathrm{g0} n_7 + \sigma_\mathrm{i}\left(n_3+n_4\right)+\sigma_\mathrm{r} n_8\right] I_\mathrm{g} \\
\frac{\mathrm{d}I_\mathrm{s}}{\mathrm{d}z} = -\sigma_s\left(n_6-n_5\right) I_\mathrm{s}
\end{dcases}
\end{align}

\noindent is integrated to obtain the pump and infrared intensities $I_\mathrm{g,out}$ and $I_\mathrm{s,out}$ after propagation through an active pixel.
If the microwaves are turned off or on, after the propagation, $I_\mathrm{s,out}\!\left(0\right)$ and $I_\mathrm{s,out}\!\left(\Omega_\mathrm{R}\right)$ describe the resulting infrared intensities, respectively.
Consequently, the ODMR contrast 

\begin{align}
C = \frac{I_\mathrm{s,out}\!\left(0\right)-I_\mathrm{s,out}\!\left(\Omega_\mathrm{R}\right)}{I_\mathrm{s,out}\!\left(0\right)}
\end{align}

\noindent is defined as the relative change in the resulting infrared intensities.

\begin{table}
\centering
\caption{\textbf{Parameters used for ODMR contrast simulations.} Refer to \citep{Dumeige2019} for the labeling of NV$^-$ and NV$^0$ energy levels. It is assumed that $T_2^*$ stays the same for diamond substrates A and B. This approximation is justified since the ODMR contrast only slightly depends on $T_2^*$ and $\Omega_\mathrm{R}$.}
\begin{tabular}{lrr}
\label{t:parameters}
Parameter & Value & Reference \\ 
\hline 
$\lambda_\mathrm{g}$ & $\SI{532}{\nano\meter}$ & \citep{Acosta2010}\\ 
$\lambda_\mathrm{s}$ & $\SI{1042}{\nano\meter}$ & \citep{Acosta2010}\\ 
$\sigma_\mathrm{g}$ & $\SI{3.1e-21}{\meter^2}$ & \citep{Wee2007} \\ 
$\sigma_\mathrm{g0}$ & $\SI{5.4e-21}{\meter^2}$ & \citep{Meirzada2018} \\ 
$\sigma_\mathrm{i}$ & $\SI{3.1e-21}{\meter^2}$ & \citep{Meirzada2018} \\ 
$\sigma_\mathrm{r}$ & $\SI{2.6e-21}{\meter^2}$ & \citep{Meirzada2018} \\ 
$\sigma_\mathrm{s}$ & $\SI{6.1e-23}{\meter^2}$ & \citep{Dumeige2019} \\ 
$\Omega_\mathrm{R}$ & $\SI{8.2e6}{\hertz}$ & estimated\\ 
$T_2^*$ & $\SI{520}{\nano\second}$ & refer to \autoref{ss:ir-power-noise}\\
$k_{31}$ & $\SI{5e7}{\hertz}$ & \citep{Bogdanov2018} \\ 
$k_{35}$ & $\SI{7.9e6}{\hertz}$ & \citep{Tetienne2012} \\ 
$k_{42}$ & $\SI{5e7}{\hertz}$ & \citep{Bogdanov2018} \\ 
$k_{45}$ & $\SI{5.3e7}{\hertz}$ & \citep{Tetienne2012} \\ 
$k_{56}$ & $\SI{1e9}{\hertz}$ & \citep{Acosta2010} \\ 
$k_{61}$ & $\SI{1e6}{\hertz}$ & \citep{Tetienne2012} \\ 
$k_{62}$ & $\SI{7e5}{\hertz}$ & \citep{Tetienne2012} \\ 
$k_{87}$ & $\SI{5.3e7}{\hertz}$ & \citep{Meirzada2018} \\ 
\end{tabular}
\end{table}

\subsection{Contrast estimation}
\label{ss:contrast-estimation}
We extract the ODMR contrast from our IRA-ODMR measurements by firstly fitting the $f_+$ resonance in an ODMR signature with the derivative of a Gaussian dip
\begin{align}
U\!\left(f_\mathrm{c}\right) = \frac{U_0\left( f_\mathrm{c} - f_+\right)}{\sqrt{2\mathrm{\pi}}\tilde{\nu}_+^3} \exp\left(\frac{-\left( f_\mathrm{c} - f_+\right)^2}{2\tilde{\nu}_+^2}\right)\,,
\end{align}
where $U_0$ is a positive constant and $\tilde{\nu}_+$ the fitted inhomogeneously broadened standard deviation linewidth.
$\tilde{\nu}_+$ relates to the full width at half maximum linewidth $\nu_+$.
Fitting with a derivative of a Gaussian dip is possible since the ODMR signal recorded with the lock-in amplifier using frequency-modulation is proportional to the ODMR dip's derivative.
Secondly, we integrate the fitted Gaussian dip's derivative sampled by $P$ samples $p,q\in\left[0,P\right)$ with $\Delta f_\mathrm{c}$ frequency spacing \citep{Stuerner2021}
\begin{align}
\hat{U}_q = \sum_{p\leq q} U_p \frac{\Delta f_\mathrm{c}}{2 f_\mathrm{dev}}\,.
\end{align}
Finding the minimum of the Gaussian dip and normalizing it with the infrared power on the balanced photodetector (refer to \autoref{ss:shot-noise}) yields the ODMR contast
\begin{align}
C= \left|\min_q \hat{U}_q\right| \frac{g_\mathrm{Mon}}{\bar{U}_\mathrm{Mon} g_\mathrm{Sig}} \,.
\end{align}
The ODMR contrast does not depend on the infrared power.
We confirm this for the single-pixel sensor in \subfigref{fig:RefIRNoise}{a}.

\subsection{Infrared power-dependent contrast and sensitivity}
\label{ss:ir-power-noise}
\begin{figure*}
	\centering
	\includegraphics[scale=1]{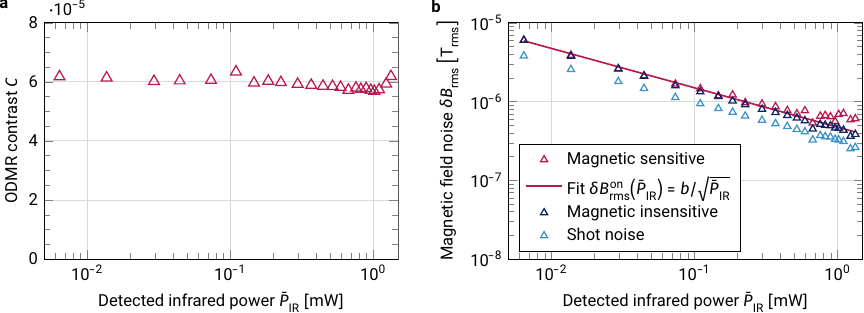}
	\caption{\textbf{Contrast and noise of the single-pixel sensor.}
	The data has been acquired under the conditions outlined in the main manuscript's methods section with a lock-in amplifier time constant $\tau_\mathrm{c}=\SI{1}{\milli\second}$.
	\subfiglabel{a} ODMR contrast $C$ and
	\subfiglabel{b} magnetic field noise $\delta B_\mathrm{rms}$ for the magnetic sensitive and insensitive case as well as for the shot noise limit depending on the infrared power $\bar{P}_\mathrm{IR}$ averaged over both photodiodes of the balanced photodetector.
	}
	\label{fig:RefIRNoise}
\end{figure*}

Identifying the ratio $\nu_+ / C$ as $\bar{U}_\mathrm{Mon} / a_+ = \bar{P}_\mathrm{IR} g_\mathrm{Mon} R / a_+$ and integrating \autoref{eq:S_SN_B} over the measurement bandwidth yields the rms photon shot noise equation given in \citep{Acosta2010} up to a system-specific factor $2 g_\mathrm{Sig} / g_\mathrm{Mon}$.
Likewise, measuring the infrared power-dependent rms noise at resonance $\delta B_\mathrm{rms}^\mathrm{on}$ allows to extract the electron spin dephasing time $T_2^*$ since with the lock-in cutoff frequency $f_\mathrm{BW}$ it is
\begin{align}
\delta B_\mathrm{rms}^\mathrm{on} = \frac{2}{\gamma} \frac{g_\mathrm{Sig}}{g_\mathrm{Mon}} \frac{\nu_+}{C} \sqrt{h \frac{c}{\lambda_\mathrm{s}} f_\mathrm{BW}} \times \frac{1}{\sqrt{\bar{P}_\mathrm{IR}}}\,,
\end{align}
where the resonance linewidth is related to the spin dephasing time according to $T_2^* = 1/\left(\mathrm{\pi}\nu_+\right)$ \citep{Dumeige2013}.
Fitting $\delta B_\mathrm{rms}^\mathrm{on}\!\left(\bar{P}_\mathrm{IR}\right) = b / \sqrt{\bar{P}_\mathrm{IR}}$ yields
\begin{align*}
T_2^* = \frac{2}{\gamma} \frac{g_\mathrm{Sig}}{g_\mathrm{Mon}} \frac{1}{\mathrm{\pi} b C} \sqrt{h \frac{c}{\lambda_\mathrm{s}} f_\mathrm{BW}}
\end{align*}
with an uncertainty derived from contrast samples plotted in \subfigref{fig:RefIRNoise}{a}
\begin{align*}
\tilde{T}_2^* = \frac{2}{\gamma} \frac{g_\mathrm{Sig}}{g_\mathrm{Mon}} \frac{1}{\mathrm{\pi} b C^2} \sqrt{h \frac{c}{\lambda_\mathrm{s}} f_\mathrm{BW}} \times \underset{\bar{P}_\mathrm{IR}}{\mathrm{std}}\!\left[C\!\left(\bar{P}_\mathrm{IR}\right)\right]\,.
\end{align*}

For the single-pixel sensor and $\bar{P}_\mathrm{IR}\lesssim\SI{0.8}{\milli\watt}$, the measured infrared power-dependent rms noise reveals a clear $1/\sqrt{\bar{P}_\mathrm{IR}}$ proportionality indicating that the sensor is mostly limited by infrared shot noise (\subfigref{fig:RefIRNoise}{b}).
This justifies the extraction of $T_2^*$ by the method introduced here.
In the $\bar{P}_\mathrm{IR}\lesssim\SI{0.8}{\milli\watt}$ regime, the magnetic sensitive and insensitive noise levels coincide.
For higher infrared powers, the magnetic sensitive noise values rise above the magnetic insensitive noise floor since another technical noise component starts to dominate.
Across the entire infrared power range, the magnetic insensitive noise floor remains slightly above the shot noise limit calculated according to \autoref{eq:S_SN_B}.
We attribute the constant offset between the magnetic insensitive noise floor and the shot noise limit to technical excess noise.

\newpage
\printbibliography

%% file: header.tex
\hyphenation{neg-a-tive-ly-charged wave-guide nano-beam}

\title{Diamond-on-chip infrared absorption magnetic field camera}

\author[1,2]{Julian~M.~Bopp}
\author[3]{Hauke~Conradi}
\author[2]{Felipe~Perona}
\author[1]{Anil~Palaci}
\author[1]{Jonas~Wollenberg}
\author[2]{Thomas~Flisgen}
\author[2]{Armin~Liero}
\author[2]{Heike~Christopher}
\author[3]{Norbert~Keil}
\author[4]{Wolfgang~Knolle}
\author[2]{Andrea~Knigge}
\author[2]{Wolfgang~Heinrich}
\author[3]{Moritz~Kleinert}
\author[1,2,*]{Tim~Schröder}
\affil[1]{Humboldt-Universität zu Berlin, Department of Physics, 12489 Berlin, Germany}
\affil[2]{Ferdinand-Braun-Institut gGmbH, Leibniz-Institut für Höchstfrequenztechnik, 12489 Berlin, Germany}
\affil[3]{Fraunhofer-Institut für Nachrichtentechnik, Heinrich-Hertz-Institut, 10587 Berlin, Germany}
\affil[4]{Leibniz-Institut für Oberflächenmodifizierung e.V., 04318 Leipzig, Germany}
\affil[*]{Corresponding author: Tim Schröder, tim.schroeder@physik.hu-berlin.de}

%% file: content.tex
\placeabstract{%
Integrated and fiber-packaged magnetic field sensors with a sensitivity sufficient to sense electric pulses propagating along nerves in life science applications and with a spatial resolution fine enough to resolve their propagation directions will trigger a tremendous step ahead not only in medical diagnostics, but in understanding neural processes.
Nitrogen-vacancy centers in diamond represent the leading platform for such sensing tasks under ambient conditions.
Current research on uniting a good sensitivity and a high spatial resolution is facilitated by scanning or imaging techniques.
However, these techniques employ moving parts or bulky microscope setups.
Despite being far developed, both approaches cannot be integrated and fiber-packaged to build a robust, adjustment-free hand-held device.
In this work, we introduce novel concepts for spatially resolved magnetic field sensing and 2-D gradiometry with an integrated magnetic field camera.
The camera is based on infrared absorption optically detected magnetic resonance (IRA-ODMR) mediated by perpendicularly intersecting infrared and pump laser beams forming a pixel matrix.
We demonstrate our 3-by-3 pixel sensor's capability to reconstruct the position of an electromagnet in space.
Furthermore, we identify routes to enhance the magnetic field camera's sensitivity and spatial resolution as required for complex sensing applications.
}

\noindent Solid-state magnetic field sensors offer outstanding sensitivities at room temperature.
Particularly, sensing with negatively-charged nitrogen-vacancy (NV) centers in diamond has evolved into a well-established technique not only limited to material sciences \citep{Clevenson2015, Chatzidrosos2019, Stefan2021, Zhou2021} and biological applications \citep{Sage2013, Barry2016, Fescenko2019, Webb2021, Arai2022, Aslam2023, Hansen2023}.
This is the case since operating NV centers under bias magnetic fields up to several Tesla is feasible and since the involved measurement sensitivities are approaching fundamental quantum limits \citep{Wolf2015, Zhang2021}.

Nowadays, there are three different NV-based sensor types.
Each type is suitable for certain applications, but none of them combines the advantages of fully integrated photonic devices and multi-pixel imaging sensors. 
Firstly, there are miniaturized single-pixel hand-held devices \citep{Webb2019, Kuwahata2020, Patel2020, Stuerner2021, Xie2022, Graham2023}.
Secondly, scanning magnetometers provide highest spatial resolutions by scanning a diamond tip across a sample \citep{Maletinsky2012, Stefan2021, Scheidegger2022, Welter2022}.
However, their moving, and thus sensitive parts prevent scanning magnetometers from becoming mobile devices by on-chip integration.
Thirdly, approaches utilizing lock-in amplifier cameras are capable of recording entire magnetic field images with sub-millisecond temporal resolution without the need for any moving components.
Though, lock-in camera approaches suffer from limited per-pixel sensitivity \citep{Parashar2022, Webb2022}.
Photonic integration of lock-in camera magnetometers is hardly possible due to the involved bulky optical components like a microscope objective.
Consequently, full photonic integration remains an obstacle for next-generation NV magnetic field imagers suitable for neuroscience applications \citep{Barry2016}.

In our work, we bridge the gap between single-pixel hand-held sensors and complex imaging magnetometers involving either sensitive moving parts or bulky optics.
We propose and demonstrate an integrated magnetic field camera capable of measuring magnetic fields with pixels aligned as a two-dimensional matrix in a diamond substrate.
Being fiber-packaged, the camera does neither require any free-space optical components nor a strong pump laser beam illuminating a sample from the top, which might damage biological samples.
In contrast to the vast majority of NV-based magnetic field sensors, we exploit the NV$^-$ $^{1}\!A_{1}\leftrightarrow{} ^{1}\!E$ singlet transition, which emits and absorbs $\SI{1042}{\nano\meter}$ infrared light upon $\SI{532}{\nano\meter}$ pumping if the NV$^-$'s electron spin is flipped to the $m_\mathrm{s}=\pm 1$ state \citep{Rogers2008, Acosta2010a}.
This infrared absorption optically detected magnetic resonance (IRA-ODMR) technique opens up entirely new routes to NV magnetometry despite its to date inferior measurement sensitivity compared to red fluorescence ODMR \citep{Acosta2010, Jensen2014, Chatzidrosos2017}.
In this way, IRA-ODMR merged with our novel sensor design allows for spatially defining magnetic field sensitive pixels: an array of collimated green pump laser beams intersects perpendicularly with an array of collimated infrared laser beams (\subfigref{fig:Pixels}{a}).
Turning on the respective pump beam in row $i\in\left[1,N\right]$ and the infrared beam in column $j\in\left[1,N\right]$ enables selective addressing and readout of the pixel $\left( i,j\right)$ within the sensor matrix.
To reconstruct an image of the magnetic field across the sensor, the pixels are subsequently selected and read out using an IRA-ODMR measurement.
Such images intrinsically contain information on a two-dimensional magnetic field gradient applied to the sensor.

\section*{A multi-pixel magnetic field camera}
\begin{figure*}
	\centering
	\includegraphics[scale=1]{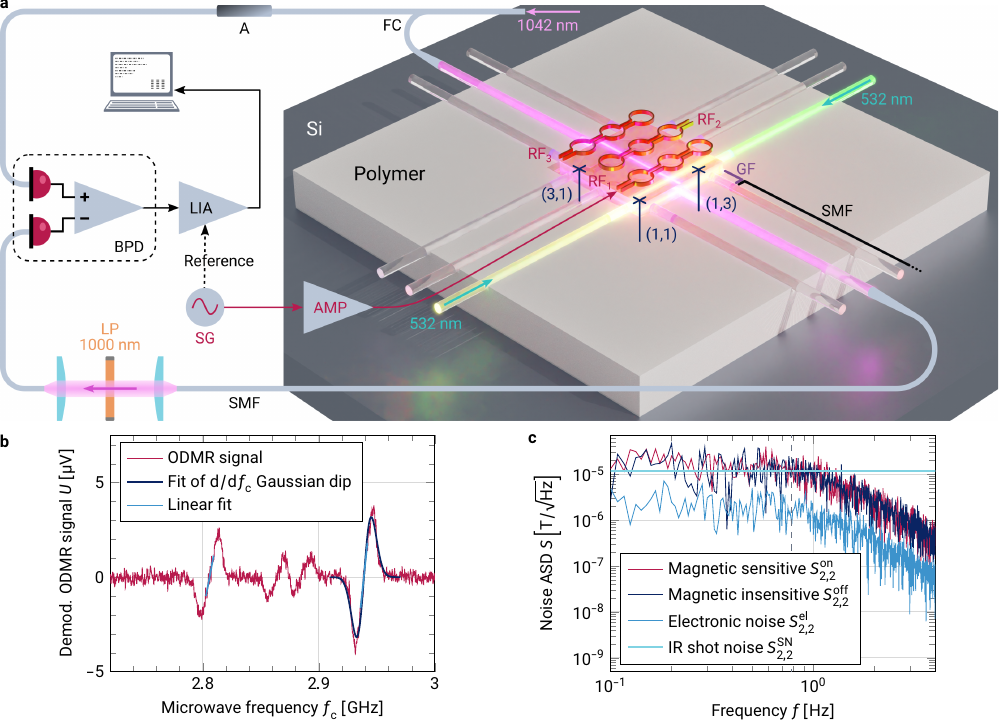}
	\caption{\textbf{Working principle of the chip-integrated NV-based multi-pixel magnetic field camera.}
	\subfiglabel{a} A diamond substrate and single-mode fibers (SMF) terminated by beam-collimating graded-index fibers (GF) are glued into trenches etched into a polymer substrate.
	 The polymer covers a silicon submount (Si) for stability.
	 Intersecting $\SI{532}{\nano\meter}$ pump and $\SI{1042}{\nano\meter}$ infrared laser beams define the magnetic sensitive pixel $\left( i,j\right)$ inside the diamond substrate. $\num{10}\,\%$ of the injected infrared light is split off by a fiber coupler (FC) and directed to the reference input of a balanced photodetector (BPD) after passing an adjustable fiber attenuator (A). The infrared light transmitted through the diamond substrate passes a $\SI{1000}{\nano\meter}$ longpass filter (LP) located in a fiber U-bench before it reaches the BPD's signal input. The BPD's difference signal is amplified by a lock-in amplifier (LIA) and processed with a computer. A signal generator (SG) provides a reference for the LIA-demodulation and generates a frequency-modulated microwave signal which is amplified (AMP) and fed into the respective microwave inductor line (RF$_i$) located above the row of active (pumped) pixels.
	 The full mechanical setup is detailed in supplementary information A.1.
	\subfiglabel{b} Exemplary IRA-ODMR signature (red curve) of pixel $\left(\num{2},\num{2}\right)$ with an offset magnetic field.
	The light blue curves represent linear fits around the two resonances used for sensing.
	One resonance is fitted by a derivative of a Gaussian function (dark blue curve).
	\subfiglabel{c} Noise amplitude spectral densities $S$ related to the $\SI{2.939}{\giga\hertz}$ resonance displayed in \subfiglabel{b} recorded on-resonance (magnetic sensitive), off-resonance (magnetic insensitive), and with the infrared beam turned off (electronic noise). The cyan line visualizes the infrared optical shot noise imposed on the signal detected by the BPD. The gray dashed line indicates the lock-in cutoff frequency.
	}
	\label{fig:Pixels}
\end{figure*}

To enable IRA-ODMR by driving the NV$^-$'s electron spin, we place microwave inductor lines RF$_i$ on top of the diamond substrate (\subfigref{fig:Pixels}{a}).
We modulate the time $t$-dependent spin-driving microwave electromagnetic field's frequency according to
\begin{equation}
f_\mathrm{RF}\!\left(t\right)=f_\mathrm{c} + f_\mathrm{dev} \cos\!\left(2\mathrm{\pi} f_\mathrm{mod} t\right) \label{eq:f_RF}
\end{equation}
around a center frequency $f_\mathrm{c}$ with a modulation frequency $f_\mathrm{mod}$ and a modulation depth $f_\mathrm{dev}$.
The center frequency-dependent difference signal of a balanced photodetector proportional to the infrared absorption is demodulated with a lock-in amplifier and recorded as the ODMR signal $U\!\left(f_\mathrm{c}\right)$ (refer to methods section).
By sweeping $f_\mathrm{c}$, we first perform IRA-ODMR measurements on each pixel to verify their integrity.
Without any offset magnetic field applied, each pixel reveals a clear ODMR signature in $U\!\left(f_\mathrm{c}\right)$ around $\SI{2.87}{\giga\hertz}$.
For pixel $\left(\num{1},\num{1}\right)$, the ODMR signature is very weak due to a reduced local overlap between intersecting pump and infrared beams.
All ODMR signatures vanish turning either the pump or the infrared beams off (refer to supplementary information B.1).
This unambiguously proves that the recorded ODMR signatures originate from infrared absorption.

To render our sensor magnetic field sensitive, we split the four possible NV orientations into two ensembles by applying an offset magnetic field aligned along the $\left[0 \bar{1} 1\right]$-direction:
The $\left[1 1 1\right]$- and $\left[1 \bar{1} \bar{1}\right]$-oriented subensembles constitute ensemble $\num{1}$.
They do not respond to the offset magnetic field in first approximation.
Whereas, the $\left[\bar{1} 1 \bar{1}\right]$- and $\left[\bar{1} \bar{1} 1\right]$-oriented subensembles (ensemble $\num{2}$) are split by the same frequency of about $\SI{133}{\giga\hertz}$.
\subfigref{fig:Pixels}{b} shows a resonance-split ODMR signature exemplarily for the center pixel $\left(\num{2},\num{2}\right)$.
Data for the other pixels is displayed in supplementary information B.2.
The pixel $\left(\num{1},\num{1}\right)$ is excluded from the following analysis for an insufficient signal-to-noise ratio.
For each of the two ensembles, there are two resonances caused by the two $m_\mathrm{s}=0 \rightarrow\pm 1$ transitions.
We attribute the different magnitudes of the four visible resonances to the microwave polarization:
the microwave magnetic field produced by a ring-shaped inductor stacked on top of the diamond substrate's $\left(111\right)$-oriented top surface is parallel to the $\left[1 1 1\right]$-oriented NV centers (and perpendicular to the offset magnetic field), thus reducing the magnitude of the first ensemble's resonances \citep{Muenzhuber2020}.
To determine the lower $f_-$ and higher $f_+$ resonance frequencies of the NV ensemble $\num{2}$ used for sensing as well as the corresponding slopes $a_\pm$, we perform linear fits $U\!\left(f_\mathrm{c}\right)=a_\pm \left(f_\mathrm{c}-f_\pm\right)$ around the $m_\mathrm{s}=0 \rightarrow\pm 1$ resonances.
An additional fit of the $f_+$ resonance by the derivative of a Gaussian function yields the pixel-averaged inhomogeneously broadened linewidth as $\SI{6.4(3)}{\mega\hertz}$ standard deviation.
Inhomogeneous broadening occurs due to the high NV density and $f_\mathrm{dev}$ spanning all hyperfine transitions \citep{ElElla2017}.
Despite causing line broadening, such a high modulation depth $f_\mathrm{dev}$ is necessary to achieve reasonable signal-to-noise ratios in the ODMR measurements.
Integrating the Gaussian-type fit leads to the ODMR contrast $C$ \citep{Dumeige2013} after normalization with the detected infrared power (refer to supplementary information D.2).
For pixel $\left(\num{3},\num{3}\right)$, we obtain the highest contrast $C_{\num{3},\num{3}}=\num{5.1e-6}$, for pixel $\left(\num{3},\num{2}\right)$ the lowest contrast $C_{\num{3},\num{2}}=\num{0.5e-6}$, and for the pixel-average $\bar{C}=\num{23(16)e-7}$.
The observed values agree with the contrast of $\num{2.3e-06}$ obtained by modeling the NV including its neutral charge state \citep{Dumeige2019} (refer to supplementary information D.1).
Non-optimal pump and infrared beam overlaps may reduce the measured contrast compared to the simulated value.
Increased values might originate from uncertainties in measuring the infrared power impinging on the balanced photodetector and from uncertainties in the transition rates and absorption cross sections employed to model the contrast.

The frequency-dependent magnetic field sensitivity of each pixel is best characterized by the amplitude spectral densities (ASD) of different noise contributions as depicted in \subfigref{fig:Pixels}{c}.
Magnetic sensitive (insensitive) noise ASDs $S^\mathrm{on}$ $\left( S^\mathrm{off}\right)$ are derived using a Fourier transform of a $\SI{100}{\second}$-long time series of the demodulated ODMR signal $U$ recorded at $f_\mathrm{c}=f_+$ $\left(f_\mathrm{c}=\SI{2.98}{\giga\hertz}\right)$ as detailed in supplementary information C.1.
Up to the lock-in cutoff frequency, both $S^\mathrm{on}$ and $S^\mathrm{off}$ show a coinciding flat plateau which closely matches the ASD of optical shot noise $S^\mathrm{SN}$ caused by infrared photons impinging on the balanced photodetector (refer to supplementary information C.2).
Contrarily, the electronic noise floor $S^\mathrm{el}$ recorded at $f_\mathrm{c}=f_+$ with the infrared beam turned off remains at lower values compared to $S^\mathrm{SN}$ emphasizing that our sensor is currently limited by the low intensity of the detected infrared signal. 
In accordance with the contrast estimation, pixel $\left(\num{3},\num{3}\right)$ reveals the best sensitivity of $\SI{10.6}{\micro\tesla}_\mathrm{rms}$ and pixel $\left(\num{3},\num{2}\right)$ the worst sensitivity of $\SI{44.0}{\micro\tesla}_\mathrm{rms}$ by integrating the respective $S^\mathrm{on}_{i,j}$ up to the lock-in cutoff frequency.
Thus, the pixel-averaged sensitivity becomes $\SI{25(10)}{\micro\tesla}_\mathrm{rms}$ (refer to supplementary information B.2).

\subsection*{Magnetic field imaging}
After confirming the pixels' integrity, we now apply the magnetic field camera to capture an image of the magnetic field generated by a current-driven solenoid coil with $N_\mathrm{s}=\num{15}$ windings, a radius of $R_\mathrm{s}=\SI{450}{\micro\meter}$, and a length of $L_\mathrm{s}=\SI{6.1}{\milli\meter}$.
The solenoid is located above pixel $\left(\num{3},\num{3}\right)$ on top of the diamond substrate.
The magnetic field to be probed points along the $\left[1 1 1\right]$-direction.
Hence, it affects both the $\left[\bar{1} 1 \bar{1}\right]$- and $\left[\bar{1} \bar{1} 1\right]$-oriented NV subensembles (ensemble $\num{2}$) in the same way.

\begin{figure*}
	\centering
	\includegraphics[scale=1]{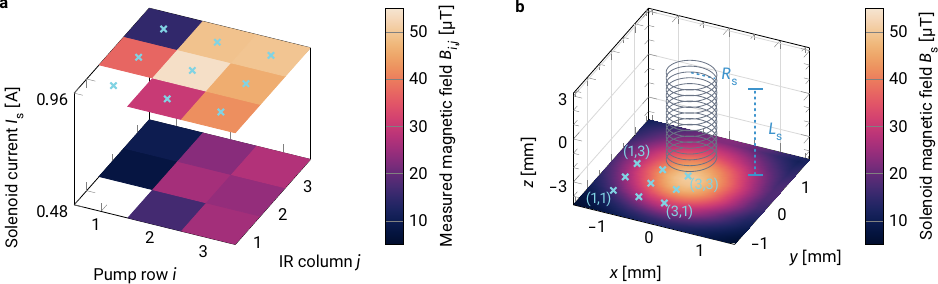}
	\caption{\textbf{Magnetic field imaging for magnetic object detection.}
	\subfiglabel{a} A solenoid coil driven by a current $I_\mathrm{s}$ is placed above pixel $\left(\num{3},\num{3}\right)$. For two different currents, the coil produces a magnetic field, which is spatially resolved by the magnetic field camera for the pixels $\left( i,j\right)$. The measured magnetic field $B_{i,j}$ is shown color-coded. Readout from pixel $\left(\num{1},\num{1}\right)$ is not possible due to insufficient local pump and infrared beam overlap. Supplementary information B.2 presents the magnetic field uncertainty extracted for each pixel and each solenoid current as well as each pixels' sensitivity. Blue crosses denote pixel values employed for retrieving the camera position relative to the solenoid coil as depicted in \subfiglabel{b}.
	\subfiglabel{b} Schematic drawing of the solenoid coil (gray spiral) with radius $R_\mathrm{s}$, length $L_\mathrm{s}$ and the absolute value of the magnetic field $B_\mathrm{s}$ it produces for $I_\mathrm{s}=\SI{0.96}{\ampere}$ at a vertical position of $z=\SI{-4.53}{\milli\meter}$, where the magnetic field camera was localized by a minimization procedure applied to the data visualized in \subfiglabel{a}. Blue crosses indicate the retrieved locations of the camera pixels $\left( i,j\right)$.
	}
	\label{fig:Camera}
\end{figure*}

\subfigref{fig:Camera}{a} depicts the magnetic field measured for each pixel for two different solenoid currents $I_\mathrm{s}$ (refer to methods section).
The measured magnetic field increases for each pixel at the higher current.
Furthermore, for both currents, the magnetic field is strongest underneath the solenoid coil's center around pixel $\left(\num{3},\num{3}\right)$.
Moving away from pixel $\left(\num{3},\num{3}\right)$, the magnetic field decreases, which complies within sensitivity and uncertainty bounds (refer to supplementary information B.2) with a simulation of the solenoid's magnetic field as shown color-coded in \subfigref{fig:Camera}{b}.

We use the higher current $\left( I_\mathrm{s}=\SI{963}{\milli\ampere}\right)$ magnetic field image and the solenoid's simulated magnetic field to reconstruct the position of the camera relative to the solenoid.
Minimizing the position-dependent difference in the absolute values of the measured $B_{i,j}$ and simulated $B_\mathrm{s}\!\left(x,y,z\right)$ magnetic fields summed over all pixels (refer to supplementary information B.3) yields the position of the magnetic field camera.
Blue crosses in \subfigref{fig:Camera}{b} indicate the retrieved camera pixel positions.
The retrieved camera position matches the experimental conditions since the solenoid coil was indeed positioned above pixel $\left(\num{3},\num{3}\right)$.
We determine the camera's position uncertainty $(\delta x,\allowbreak\delta y,\allowbreak\delta z)\allowbreak =\allowbreak (\num{30},\allowbreak\num{20},\allowbreak\num{10})\,\si{\micro\meter}$ by Monte Carlo sampling of each pixel's magnetic field normal distribution spanned by the measured mean values $B_{i,j}$ and their uncertainties $\tilde{B}_{i,j}$.

\subsection*{Single-pixel benchmarking}
\begin{figure*}
	\centering
	\includegraphics[scale=1]{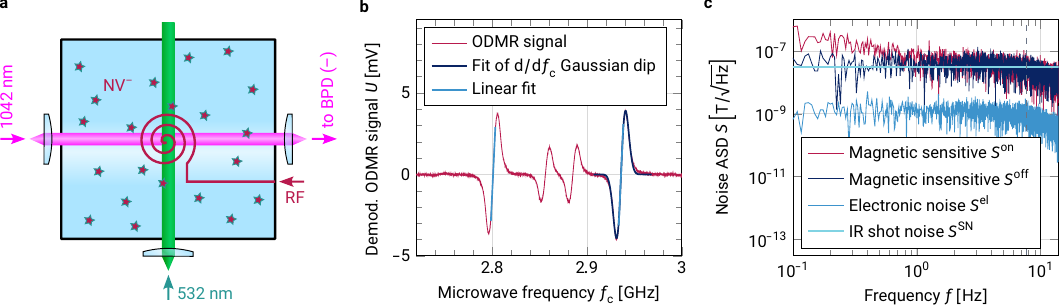}
	\caption{\textbf{Single-pixel reference measurements.}
	\subfiglabel{a} A free-space green pump beam intersects with a free-space infrared beam within a bulk diamond substrate with $\left(111\right)$-oriented top surface. A copper wire coil is located directly above the intersection to drive the NV$^-$'s spin transitions for IRA-ODMR measurements. The RF input as well as the laser inputs and outputs are connected to the measurement system as depicted in \subfigref{fig:Pixels}{a} replacing the integrated sensor but maintaining the same pixel volume as defined by the beam waists (refer to methods section).
	\subfiglabel{b} Corresponding IRA-ODMR signature (red curve) with an offset magnetic field in $\left[0 \bar{1} 1\right]$-direction. 
	\subfiglabel{c} Noise amplitude spectral densities $S$ related to the $\SI{2.936}{\giga\hertz}$ resonance displayed in \subfiglabel{b}.
	In the magnetic sensitive case, the sensitivity is $\SI{390}{\nano\tesla}_\mathrm{rms}$.
	The curves shown in \subfiglabel{b} and \subfiglabel{c} match the description in the caption of \subfigref{fig:Pixels}{b} and \subfigref{fig:Pixels}{c}, respectively.
	}
	\label{fig:Ref}
\end{figure*}

We successfully captured an image of a solenoid coil's spatial magnetic field distribution despite the sensitivity limitations of our multi-pixel magnetic field camera.
These limitations originate from (i) the non-optimal pump and infrared beam overlap within the pixel volumes, (ii) restricted laser powers due to the adhesive's damage threshold, and (iii) the weak interaction cross section between infrared photons and the NV$^-$ singlet transition without cavity-enhancement \citep{Dumeige2019}.
Whereas overcoming constraint (iii) is not straightforward without employing an optical cavity for infrared photons, we now examine a single-pixel sensor which relieves constraints (i) and (ii).
In a setup utilizing free-space optics, the pump and infrared beams are adjusted to intersect perfectly within a second diamond substrate (\subfigref{fig:Ref}{a}).
A similar infrared power but an at least $\num{64}$ times higher pump power compared to the multi-pixel sensor is applied (refer to methods section).
\subfigref{fig:Ref}{b} depicts the corresponding IRA-ODMR signature.
Besides the expected increase in the signal-to-noise ratio under optimized measurement conditions, the ODMR signature resembles the multi-pixel sensor's ODMR signatures closely (compare with \subfigref{fig:Pixels}{c}). 
The upper resonance at $f_+=\SI{2.936}{\giga\hertz}$ features a $\SI{4.8}{\mega\hertz}$ inhomogeneously broadened linewidth (standard deviation) as well as an ODMR contrast of $C=\num{6.6e-5}$.
The difference between the contrast of $\num{1.7e-5}$ as modeled for the single-pixel sensor configuration according to \citep{Dumeige2019} (refer to supplementary information D.1) and the measured contrast hints at the need to determine the NV transition rates and absorption cross sections more precisely.

Last, we find significantly better sensitivities as shown in \subfigref{fig:Ref}{c}.
Within the $\num{10}$ times larger measurement bandwidth compared to the multi-pixel sensor, the sensitivity improves to $\SI{390}{\nano\tesla}_\mathrm{rms}$ in the magnetic sensitive case.
While the magnetic insensitive amplitude spectral density is completely flat up to the lock-in cutoff frequency, there is some excess noise below about $\SI{0.8}{\hertz}$ in the magnetic sensitive case raising the magnetic sensitive amplitude spectral noise density above the magnetic insensitive one.
We attribute this excess noise to low-frequency technical noise in our system.
Particularly, we do not compensate for intensity fluctuations in the pump light.
An infrared power $P_\mathrm{IR}$-dependent measurement of the magnetic sensitive rms noise $\delta B^\mathrm{on}_\mathrm{rms}$ reveals a dependency $\delta B^\mathrm{on}_\mathrm{rms}\sim 1/\sqrt{P_\mathrm{IR}}$, which allows to derive the spin dephasing time $T_2^*=\SI{520(110)}{\nano\second}$ (refer to supplementary information D.3).
Hence, the single-pixel sensor still operates in a nearly shot noise-limited regime.

The measured single-pixel sensor's sensitivity indicates the multi-pixel sensor's potential when addressing solely the above-mentioned issues (i) and (ii).
The single-pixel sensor avoids optical losses which exist in the multi-pixel case where infrared light is coupled from a fiber into the diamond substrate and collected at the opposite side facet.
These losses reduce the amount of detected infrared photons, and thus raise the relative shot noise contribution emphasizing the relevance of reducing optical losses in the infrared path through the diamond substrate.
For the multi-pixel sensor, the increased relative infrared shot noise as well as further frequency-independent electronic noise contributions obscure the technical excess noise at lower frequencies causing the magnetic sensitive and insensitive amplitude spectral densities to coincide (\subfigref{fig:Pixels}{c}).

\subsection*{Time-varying magnetic fields}
\begin{figure}
	\centering
	\includegraphics[scale=1]{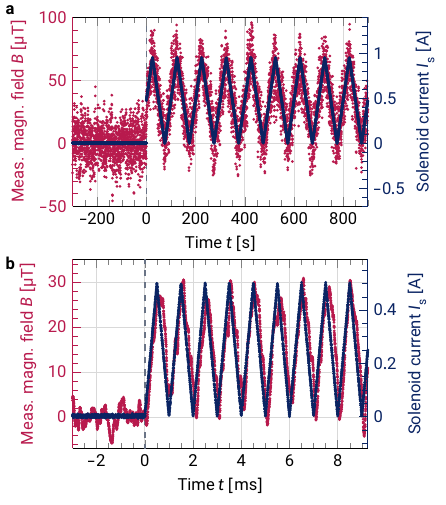}
	\caption{\textbf{Time-dependent magnetic field measurements}
	\subfiglabel{a} for the central $\left(\num{2},\num{2}\right)$ pixel of the magnetic field camera and for
	\subfiglabel{b} the single-pixel reference setup.
	The measurements in \subfiglabel{a} (\subfiglabel{b}) are performed at resonance $f_\mathrm{c}=\SI{2.939}{\giga\hertz}$ $\left(f_\mathrm{c}=\SI{2.936}{\giga\hertz}\right)$ and belong to the ODMR signatures displayed in \subfigref{fig:Pixels}{b} (\subfigref{fig:Ref}{b}).
	Red plots show the measured magnetic fields for currents $I_\mathrm{s}$ applied to a solenoid coil put on top of the respective sensor in close proximity to the active pixel.
	The periodic triangular current modulations (blue curves) start at $t=\SI{0}{\second}$ as indicated by the dashed lines and possess frequencies of $f_\mathrm{s}=\SI{10}{\milli\hertz}$ and $f_\mathrm{s}=\SI{1}{\kilo\hertz}$ in \subfiglabel{a} and \subfiglabel{b}, respectively.
	}
	\label{fig:Time}
\end{figure}

After proving the multi-pixel sensor's imaging capabilities as well as determining the single-pixel sensor's sensitivity for IRA-ODMR, we now investigate how the multi-pixel sensor responds to time-varying magnetic fields produced by current-driven solenoid coils (\subfigref{fig:Time}{a}).
The solenoid coil already employed for magnetic field imaging (\subfigref{fig:Camera}{b}) is moved directly above the multi-pixel sensor's center pixel $\left(\num{2},\num{2}\right)$.
Acquiring one ODMR signature and fixing the microwave center frequency to the upper resonance frequency enables time-dependent magnetic field measurements (refer to methods section).
When the solenoid current $I_\mathrm{s}$ is ramped up and down with a frequency $f_\mathrm{s}=\SI{10}{\milli\hertz}$, the measured magnetic field follows the current directly.
Before the modulation starts ($t<\SI{0}{\second}$), the field fluctuates around $\SI{0}{\tesla}$ with a standard deviation of $\SI{12.9}{\micro\tesla}_\mathrm{rms}$.
Corresponding to the magnetic sensitive rms noise, this value agrees well with the one extracted from data displayed in \subfigref{fig:Pixels}{c}.
In comparison to the magnetic field modulation amplitude of about $\SI{80}{\micro\tesla}$, it suggests that a higher time resolution is achievable to the disadvantage of reducing the signal-to-noise ratio.

The multi-pixel sensor's bandwidth is currently limited by its moderate sensitivity and the associated need for long averaging times.
For benchmarking, we repeat the same measurement to demonstrate that already the single-pixel reference setup's sensitivity suffices to temporally resolve a ramp signal with a frequency $f_\mathrm{s}=\SI{1}{\kilo\hertz}$ (\subfigref{fig:Time}{b}).
The measured magnetic field preserves the triangular shape of the solenoid current illustrating that the single-pixel sensor's bandwidth exceeds the ramp signal's higher frequency components.
However, the measured field is slightly delayed with respect to the applied current by the lock-in amplifier's filtering step.
For $t<\SI{0}{\second}$, we extract a magnetic sensitive noise floor of $\SI{2.2}{\micro\tesla}_\mathrm{rms}$.
Due to the increased bandwidth, this noise floor is slightly worse compared to the aforementioned optimized single-pixel sensor's sensitivity.

\section*{Towards real-time imaging of neural signals}
We have demonstrated an integrated, fiber-packaged magnetic field camera.
Our single-pixel measurements show the possibility to achieve measurement bandwidths required to sense neural signals with a duration around the millisecond regime \citep{Barry2016}.
Multiple technical improvements will lead to sensitivity enhancements and finally to a magnetic field camera for recording neural signals in real-time:
(i) Higher laser powers facilitated by non-absorbing adhesives and polymer-integrated components enhance the pixels' sensitivity as reinforced with the single-pixel sensor benchmarks.
(ii) The overlap between the pump and the infrared beams within the pixel volumes has to be optimal.
Our magnetic field camera suffers from a reduced overlap due to suboptimal polishing of the diamond side facets, which are not entirely parallel.
(iii) Local electron irradiation only within the pixel volumes allows for way higher NV densities compared to a homogeneously irradiated diamond substrate.
In this case, the pump light absorption occurs solely at the pixels.
(iv) Incorporating cavities with finesse $F$ into the infrared path enhances the contrast by a factor $2F/\mathrm{\pi}$ \citep{Jensen2014}.
Such cavities might be fabricated by coating the diamond substrate's facets.
(v) Magnetic flux concentrators amplify external magnetic fields by at least two orders of magnitude \citep{Fescenko2020}.
Altogether, the integrated magnetic field camera's fundamental spin-projection noise limit \citep{Budker2007} of about $\SI{200}{\femto\tesla\,\hertz^{-1/2}}$ certainly becomes approachable, whereas already sensitivities of about $\SI{100}{\pico\tesla\,\hertz^{-1/2}}$ enable the detection of biological signals \citep{Webb2021, Clevenson2015}.

\begin{figure}
	\centering
	\includegraphics[scale=1]{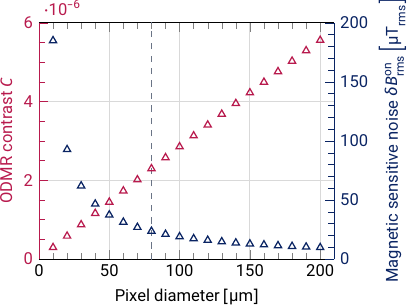}
	\caption{\textbf{Contrast (red) and sensitivity (blue) dependence on camera pixel diameter.}
	The depicted data assumes the $\left(\num{3},\num{3}\right)$ camera pixel's properties and that they, except the ODMR contrast $C$ and the magnetic sensitive rms noise $\delta B_\mathrm{rms}^\mathrm{on}$, remain constant for changing pixel diameters.
	The dashed line indicates the magnetic field camera's pixel size from this work.
	}
	\label{fig:PixelVolume}
\end{figure}

The camera's spatial resolution depends on the pixel density, which is limited by the diameter of the fibers attached to the diamond side facets.
In this proof of principle, the pixel spacing is $\SI{500}{\micro\meter}$.
Ultimately, the Rayleigh length restricts the beam diameters inside the diamond substrate to an order of magnitude of about $\SI{10}{\micro\meter}$.
However, reducing the beam diameters to obtain smaller pixels also reduces the volume where the pump and the infrared beams interact with the NV centers.
Likewise, the ODMR contrast is reduced.
To quantify the effect of the camera pixels' diameter on the ODMR contrast $C$ and on the magnetic sensitive noise, we exemplarily consider camera pixel $\left(\num{3},\num{3}\right)$.
We model its ODMR contrast for different pixel diameters to firstly rescale its ODMR signature slope $a_+$ and secondly its magnetic sensitive rms noise $\delta B_\mathrm{rms}^\mathrm{on}$ (refer to supplementary information D.2, D.3, and C.1).
While the ODMR contrast linearly rises with increasing pixel diameter, the noise rapidly drops following its $\delta B_\mathrm{rms}\sim 1/a_+$ dependency (\autoref{fig:PixelVolume}).
In this work, the pump and infrared beam waists set the camera pixels' diameter to $D_\mathrm{px} = \SI{80}{\micro\meter}$.
Reducing the diameter to its half doubles the noise.
Increasing the diameter instead does not significantly improve the pixel sensitivity.
The contrast related to $D_\mathrm{px}$ deviates slightly from the measured $C_{\num{3},\num{3}}$ due to uncertainties in the transition rates and absorption cross sections employed for contrast modeling.

Constraints on the spatial resolution as well as on the laser powers imposed by the adhesive to attach the components to the polymer can be relieved by further integrating laser diodes and photodetectors on-chip, directly next to the diamond facets.
For larger $N\times N$ pixel matrices and with prospective CMOS integration \citep{Kim2019}, more sophisticated readout schemes enable faster magnetic field image acquisition.
The lock-in demodulation technique renders simultaneous readout of multiple pixels with a single photodetector but multiple lock-in amplifier channels possible if the pixels' ODMR signals are modulated at distinct frequencies $f_\mathrm{mod}$.
We emphasize that the multi-pixel magnetic field camera can also act as a multi-pixel temperature sensor by slightly adapting the measurement procedure \citep{Kucsko2013}.
Moreover, we suggest to fabricate the microwave inductors on top of the polymer platform, i.e., directly underneath the diamond substrate in a future design iteration of the magnetic field camera.
Doing so exposes the diamond surface to samples, which can be put directly on top of the diamond substrate.
Besides the desired close proximity between sensor and sample, the chemically inert and non-toxic diamond surface is compatible with biological samples as well as with the investigation of chemically aggressive substances.
Consequently, we envision our integrated and fiber-packaged magnetic field camera to be applied in a broad range of fields:
regarding the lifetime and quality assessment of lithium-ion batteries required for the present energy revolution \citep{Bason2022}, our sensor is already able to spatially resolve magnetic field changes of several $\si{\micro\tesla}$.
Furthermore, we foresee applications in microfluidic lab-on-a-chip devices for drug discovery \citep{Allert2022} and in spaceborne quantum magnetometry \citep{Deans2023}.

\section*{Methods}
\subsection*{Multi-pixel sensor}
To realize our magnetic field camera with $\num{3}\times\num{3}$ pixels, we firstly irradiate a $\num{1.4}\times\num{1.4}\times\SI{0.2}{\milli\meter^3}$ large, $\left(111\right)$-oriented CD1411 HPHT diamond substrate A from Sumitomo Electric Industries with an electron beam and anneal it to increase the NV density to $n_\mathrm{A} = \SI{3.0e23}{\meter^{-3}}$ (refer to supplementary information A.2).
Secondly, the substrate's four side facets are polished by Delaware Diamond Knives.
Last, the diamond substrate as well as three single-mode fibers at either of the four side facets are integrated into trenches etched into a polymer board (\subfigref{fig:Pixels}{a}) and fixed with an UV-curing adhesive (refer to supplementary information A.3).
A graded-index (GRIN) fiber terminates each single-mode fiber to collimate the pump and infrared beams injected into the diamond and to couple the infrared beams back into the fibers attached to the opposite facet.
Inside the diamond substrate, both the pump and infrared beams possess beam waists of approx. $\SI{40}{\micro\meter}$.
Collimating the beams allows for infrared transmission measurements across the diamond substrate and defines the pixels as the intersections of pump and infrared beams.

Due to the NV$^-$ (NV$^0$) pump light absorption cross section of $\sigma_\mathrm{g}=\SI{3.1e-21}{\meter^2}$ \citep{Wee2007} ($\sigma_\mathrm{g0} = \num{1.8}\sigma_\mathrm{g}$ \citep{Meirzada2018}), only about $\num{19}\,\%$ of the pump light injected into the fiber at the first pixel of a row reaches the last pixel of the same row (with $\num{70}\,\%$ of the NV centers being ionized in the steady state, refer to supplementary information D.1).
To equalize the pump intensity at every pixel within a row, we inject approx. $\SI{18}{\milli\watt}$ pump light into either of the two fibers defining a row.

The infrared transmission measurement is based on a balanced detection scheme for common mode rejection (CMR) of laser intensity fluctuations superimposing the detected signal \citep{Webb2019}:
a 90:10 fiber coupler splits $\num{10}\,\%$ of the infrared light off and directs this small fraction to the positive input (BPD+) of a balanced photodetector (PDB450C from Thorlabs with $\times 10^5$ difference output gain).
The large infrared light fraction passes through the diamond and through a $\SI{1000}{\nano\meter}$ longpass filter to remove remaining red fluorescence light before it arrives at the negative input (BPD--) of the balanced photodetector.
An adjustable fiber attenuator in the infrared path bypassing the diamond substrate balances the detected intensities.
Hence, the balanced photodetector's difference signal vanishes without any ODMR-related infrared absorption.
After the $\num{90}\,\%$ output port of the fiber coupler and before the fiber interfacing the diamond substrate, we measure an infrared power of about $P_\mathrm{IR}=\SI{15}{\milli\watt}$.

Driving spin transitions to perform ODMR requires microwave magnetic fields.
For compatibility with full photonic integration, we use three impedance-matched inductor lines RF$_i$ consisting of three connected rings each fabricated onto an AlN printed circuit board (refer to supplementary information A.4).
The AlN board is aligned in close proximity to the diamond substrate's surface such that the three inductor lines lie parallel to the pump laser paths and that the pixels are centered with respect to the corresponding inductor ring.
A Rohde \& Schwarz SMB100B signal generator connected to a Mini-Circuits ZHL-16W-43-S+ amplifier and eventually to the inductor line belonging to the active pump row produces the driving microwave electromagnetic field with a time $t$-dependent frequency $f_\mathrm{RF}\!\left(t\right)$ and a power of $\SI{16}{\watt}$.
A JCC2300T3600S2R1 circulator from JQL Electronic in between the amplifier and the inductor protects the signal generator and the amplifier from reflected microwaves.
Employing the frequency modulation ODMR measurement scheme described well in \citep{Jensen2014}, we modulate the microwave signal according to \autoref{eq:f_RF} around a center frequency $f_\mathrm{c}$ with a modulation frequency $f_\mathrm{mod}=\SI{20}{\kilo\hertz}$ and a modulation depth $f_\mathrm{dev}=\SI{10}{\mega\hertz}$.
$f_\mathrm{c}$ is swept from $\SI{2.72}{\giga\hertz}$ to $\SI{3.02}{\giga\hertz}$ with a scan speed of $\SI{1}{\second\per\mega\hertz}$ and a step width of $\SI{1}{\mega\hertz}$.
$f_\mathrm{mod}$ and $f_\mathrm{dev}$ are chosen to optimize the measurement sensitivity by discarding low-frequency technical noise and by involving all NV$^-$ hyperfine transitions split by $\SI{2.16}{\mega\hertz}$ \citep{Smeltzer2009} in the frame of lock-in detection:
a Zurich Instruments MFLI lock-in amplifier demodulates the balanced photodetector's difference signal using a time constant $\tau_\mathrm{c}=\SI{100}{\milli\second}$ with a $\num{24}\,\mathrm{dB}/\mathrm{oct}$ filter roll-off.
Yielding the ODMR signal $U$, this enables the measurement of the infrared transmission's derivative.
The diamond substrate temperature is stabilized to $\SI{21}{\celsius}$ to counteract heating by pump light and microwave absorption.

For magnetic field imaging as in \subfigref{fig:Camera}{a}, we record full ODMR signatures for three solenoid currents $I_\mathrm{s}=\left\{0,482,963\right\}\,\si{\milli\ampere}$ for each pixel (refer to supplementary information B.2). We then calculate the frequency spacing $\Delta f_{i,j}\!\left(I_\mathrm{s}\right) = f_{i,j,+}\!\left(I_\mathrm{s}\right) - f_{i,j,-}\!\left(I_\mathrm{s}\right)$ between the lower and upper resonances. Subtracting the zero-current spacing yields the pixels' magnetic fields according to
\begin{align}
B_{i,j}\!\left(I_\mathrm{s}\right)=\frac{\Delta f_{i,j}\!\left(I_\mathrm{s}\right) - \Delta f_{i,j}\!\left(\num{0}\right)}{2\gamma}
\end{align}
with the NV's gyromagnetic ratio $\gamma=\SI{28}{\giga\hertz\per\tesla}$ \citep{Loubser1978}.

\subsection*{Single-pixel reference}
The single-pixel sensor's diamond substrate preparation (substrate B) equals the preparation of substrate A. 
Though, the NV density of substrate B is slightly lower $\left( n_\mathrm{B} = \SI{1.1e23}{\meter^{-3}}\right)$.
For IRA-ODMR measurements, the pump and infrared beams are collimated with free-space optics to maintain beam waists of $\SI{40}{\micro\meter}$.
Subsequently, the beams are adjusted with mirrors such that they perfectly intersect within diamond substrate B, close to its top surface and close to the side facet acting as the pump input port.
Instead of microwave inductors integrated into an AlN board, an enameled copper wire coil with a radius of $\SI{500}{\micro\meter}$, a length of $\SI{4}{\milli\meter}$, and $\num{10}$ windings located directly above the beam intersection drives spin transitions with a microwave power of $\SI{16}{\watt}$ (refer to \subfigref{fig:Ref}{a}).
Besides the sensor and the microwave inductor, measurements are performed with the setup introduced in \subfigref{fig:Pixels}{a}.
Using free-space optics instead of fibers to interface the diamond substrate, the pump laser power is risen to $\SI{1.16}{\watt}$ (measured directly in front of the diamond's pump entrance facet).
The infrared power directly in front of the diamond's respective entrance port is $\SI{9.8}{\milli\watt}$, which corresponds to $\SI{1.4}{\milli\watt}$ at either photodiode of the balanced photodetector.
Sweeping $f_\mathrm{c}$ is now performed across the same frequency interval and with the same scan speed but with a step width of $\SI{0.1}{\mega\hertz}$.
A microwave modulation frequency $f_\mathrm{mod}=\SI{87.7}{\kilo\hertz}$, a modulation depth $f_\mathrm{dev}=\SI{5}{\mega\hertz}$, and a lock-in amplifier time constant $\tau_\mathrm{c}=\SI{10}{\milli\second}$ optimize the ODMR sensitivity for the single-pixel measurement.
All other parameters remain unchanged compared to the multi-pixel sensor measurements.

\subsection*{Time-dependent measurements}
A solenoid coil stacked on top of the multi- or the single-pixel sensor generates a time-varying magnetic field being driven with a current $I_\mathrm{s}$ delivered by a Thorlabs LDC210C laser diode current controller.
For the single-pixel sensor, a Mini-Circuits VLFG-1400+ lowpass filter protects the LDC210C from microwaves coupled into the solenoid coil.
The measurements in \autoref{fig:Time} involve a current modulation according to a periodic triangular function with a minimal current of $\SI{0}{\ampere}$, a maximal current $I_\mathrm{s,max}$, and a frequency $f_\mathrm{s}$.
In case of the multi-pixel sensor, $I_\mathrm{s,max}$ is set to $\SI{958}{\milli\ampere}$, while for the single-pixel sensor, it is $I_\mathrm{s,max}=\SI{512}{\milli\ampere}$.
Recording the ODMR signal $U\!\left(t\right)$ over time $t$ at a microwave center frequency $f_\mathrm{c}=f_+\!\left(0\right)$ fixed to the upper resonance frequency for zero solenoid current and converting it with $B\!\left(t\right)=U\!\left(t\right)/\left(a_+\gamma\right)$ to a magnetic field allows for the comparison between the applied solenoid current and the measured field strength.

For the single-pixel sensor, a solenoid coil with a radius of $\SI{1.5}{\milli\meter}$, a length of $\SI{2}{\milli\meter}$, and $\num{2.5}$ windings produces the time-varying magnetic field to be gauged.
Possessing a larger diameter than the microwave coil, it is located above the intersecting pump and infrared beams surrounding the microwave coil.
Here, the lock-in amplifier's time constant is decreased to $\tau_\mathrm{c}=\SI{20}{\micro\second}$ to prevent averaging over the current ramps.
Thus, the lock-in cutoff frequency becomes $\SI{3.9}{\kilo\hertz}$ and the ODMR scan speed $\SI{20}{\milli\second\per\mega\hertz}$.

\paragraph*{Acknowledgements}
This project is funded by the German Federal Ministry of Education and Research (BMBF) within the `DiNOQuant' project (No. 13N14921) as well as by the European Research Council within the ERC Starting Grant `QUREP' (No. 851810). We acknowledge further funding from the Einstein Research Unit `Perspectives of a quantum digital transformation: Near-term quantum computational devices and quantum processors'.
The authors thank Laura Orphal-Kobin and Masazumi Fujiwara for fruitful discussions on applications of the magnetic field camera as well as Viviana Villafane for annealing the diamond substrates.

\paragraph*{Author contributions}
J.M.B., H.Co., F.P., M.K., and T.S. invented the magnetic field camera concept.
J.M.B. designed and performed the multi-pixel sensor experiments and analyzed the data.
F.P. designed the multi-pixel sensor's electronic and mechanical parts.
H.Co. and M.K. designed the multi-pixel sensor's optical parts, H.Co. assembled them.
N.K. supervised the multi-pixel sensor polymer design and manufacturing process.
A.P. performed the single-pixel experiments.
J.W. measured the diamond substrates' NV density and Rabi frequency. 
H.Ch. developed the infrared laser diode, supervised by A.K.
W.K. was responsible for the diamond substrate electron irradiation.
T.F., A.L., and W.H. developed and fabricated the microwave inductors.
T.S. supervised the project.
J.M.B., F.P., T.F., M.K., and T.S. wrote the manuscript.
All authors reviewed it.

\paragraph*{Additional information}
Supplementary information is available in the online version of the paper.
Correspondence and requests for materials should be addressed to T.S.

\paragraph*{Competing interests}
J.M.B., H.Co., F.P., M.K., and T.S. filed the patents US11719765B2 (active) and EP4099041A1.